\documentclass[12pt]{article}
\usepackage{jhep-mod}
\usepackage{bm}
\usepackage{amssymb, amsmath, amsthm}
\usepackage{mathrsfs}
\usepackage{hyperref}
\usepackage{multicol}
\usepackage{array, tabu}
\usepackage[tableposition=top]{caption}
\usepackage{tocloft}
\usepackage{enumitem}
\usepackage[utf8]{inputenc}
\usepackage{bigints}
\usepackage{appendix}
\usepackage{graphicx}
\usepackage{placeins}
\usepackage{tikz}
\usepackage{setspace}
\usepackage{cancel}
\usepackage[normalem]{ulem}
\usepackage{physics}
\usepackage{doi}
\usepackage{csquotes}
\usepackage{mathtools}
\usepackage{subcaption}
\usepackage{breakurl}
\usepackage{lmodern}
\usepackage[T1]{fontenc}
\definecolor{purple}{rgb}{1,0,1}
\definecolor{lime}{HTML}{A6CE39} 
\definecolor{darkgreen}{rgb}{.125,.5,.25}

\definecolor{lime}{HTML}{A6CE39}
\newcommand{\orcidicon}{%
	\begin{tikzpicture}
	\draw[lime, fill=lime] (0,0) 
		circle [radius=0.16] 
		node[white] {{\fontfamily{qag}\selectfont \tiny ID}};
	\draw[white, fill=white] (-0.0625,0.095) 
		circle [radius=0.007];
	\end{tikzpicture}
	\hspace{-5mm}
}
\newcommand\orcidJessica{{\href{https://orcid.org/0000-0002-2669-2899}{\orcidicon}}}
\newcommand\orcidSebastian{{\href{https://orcid.org/0000-0003-1997-0026}{\orcidicon}}}
\newcommand\orcidMatt{{\href{https://orcid.org/0000-0003-1088-6485}{\orcidicon}}}
\begin{document} 
\title{
\huge{
Tractor beams, pressor beams, \\ 
\leftline{and stressor beams in general relativity}
}
}
\author{
\Large
\leftline{Jessica Santiago$^1$\orcidJessica\!,
Sebastian Schuster$^2$\orcidSebastian\!, {\sf{and}}
Matt Visser$^1$\orcidMatt}}
\affiliation{$^1$ School of Mathematics and Statistics, Victoria University of Wellington, \\
\null\qquad PO Box 600, Wellington 6140, New Zealand.}
\affiliation{$^2$ Institute of Theoretical Physics,
Faculty of Mathematics and Physics, Charles University,\\
\null\qquad V~Hole\v{s}ovi\v{c}k\'{a}ch~2, 180 00 Prague 8, Czech Republic}

\emailAdd{jessica.santiago@sms.vuw.ac.nz}
\emailAdd{sebastian.schuster@utf.mff.cuni.cz}
\emailAdd{matt.visser@sms.vuw.ac.nz}
\parindent0pt
\parskip7pt

\abstract{

The metrics of general relativity generally fall into two categories: Those which are solutions of the Einstein equations for a given source energy-momentum tensor, and the \enquote{reverse engineered} metrics---metrics bespoke for a certain purpose. Their energy-momentum tensors are then calculated by inserting these into the Einstein equations. This latter approach has found frequent use when confronted with creative input from fiction, wormholes and warp drives being the most famous examples. In this paper, we shall again take inspiration from fiction, and see what general relativity can tell us about the possibility of a gravitationally induced tractor beam. We will base our construction on warp drives and show how versatile this ansatz alone proves to be. Not only can we easily find tractor beams (attracting objects); repulsor/pressor beams are just as attainable, and a generalization to \enquote{stressor} beams
is seen to present itself quite naturally. We show that all of these metrics would violate various energy conditions. This will provide an opportunity to ruminate on the meaning of energy conditions as such, and what we can learn about whether an arbitrarily advanced civilization might have access to such beams.

\bigskip
\noindent
{\sc Date:} Wednesday 9 June 2021; \LaTeX-ed \today

\bigskip
\noindent
{\sc Keywords}: tractor beams; pressor beams; stressor beams; modified warp drives; general relativity; metric engineering; energy conditions.
}

\maketitle

\def\H{{\scriptscriptstyle{\mathrm{H}}}}
\def\ISCO{{\scriptscriptstyle{\mathrm{ISCO}}}}
\def\d{{\mathrm{d}}}
\def\O{{\mathcal{O}}}
\def\F{{\mathcal{F}}}
\def\sign{{\mathrm{sign}}}
\def\L{{\mathcal{L}}}
\def\NEC{{\hbox{NEC}}}
\def\WEC{{\hbox{WEC}}}
\def\SEC{{\hbox{SEC}}}
\def\DEC{{\hbox{DEC}}}
\def\NCC{{\hbox{NCC}}}
\def\TCC{{\hbox{TCC}}}
\def\tr{{\mathrm{tr}}}
\def\e{{\mathrm{e}}}
\newcommand\R[1]{{\mathbb{R}^{#1}}}
\addtocontents{toc}{\protect\vspace{-15pt}}
\section{Introduction}\label{S:Intro}
\addtocontents{toc}{\protect\enlargethispage{2.5\baselineskip}}

Within the context of standard general relativity, there has now been over 33 years of serious theoretical work on the possibility of ``traversable wormholes''~\cite{Morris:1988a,Morris:1988b,Visser:1989a,Visser:1989b,Book}, 29 years of recent work  on ``time machines''~\cite{protection,hawking60,Visser:1992a,Visser:1992b,Friedman:2008,do-not-mess, tardis}, and over 27 years of work on the  theoretical possibility of ``warp drives''~\cite{Alcubierre:1994,Natario:2001,Lobo:2004,Lobo:2017,Alcubierre:2017,Santiago:2021}. These analyses, and their subsequent refinements, are based on ``reverse engineering'' the space-time metric to encapsulate some potentially interesting physics, and then using the Einstein equations to deduce what the stress-energy tensor must be to support these space-times~\cite{Morris:1988a,Book,cloak,cthulu}.  

A distinct century-old trope within science fiction is the tractor/pressor beam~\cite{tractor,pressor1,pressor2}. 
To the best of our knowledge, no really focussed work has been carried out on putting  tractor/pressor beams into a coherent general relativistic context. (Acoustic tractor beams~\cite{acoustic, acoustic2, acoustic3, acoustic4}, 
matter wave tractor beams~\cite{matter}, or optical tweezers~\cite{tweezers}, seem to be the closest one gets in the current scientific literature.) 

Herein we shall analyze tractor/pressor/stressor beams from a general relativistic perspective. 
The basic idea is to significantly modify and adapt the ``warp drive'' space-times~\cite{Alcubierre:1994,Natario:2001,Lobo:2004,Lobo:2017,Alcubierre:2017,Santiago:2021} in a suitable manner, giving them a ``beam like'' profile, and analysing the induced stresses and forces. Instead of a spaceship riding inside a warp bubble, we will assume that the warp field is in the form of a ``beam'' generated to pull/repel a target. 
The mechanisms by which this field is generated is beyond the scope of this article. We will assume that some arbitrarily advanced civilisation \cite{scale,scale2} might have developed the appropriate beam generation technology.

Specifically, we shall assume for convenience that the modified warp drive space-times are oriented in the $z$ direction and give them a uniform transverse profile in the $x$ and $y$ directions, typically of the form $f(x^2+y^2)$. Doing so, one obtains a ``beam'' rather than a ``warp bubble''. Note that in this work we will let the $(t,z)$ dependence remain arbitrary.

As always, when working in this area of speculative physics, including wormholes, and warp drives, and now  tractor/\-pressor/\-stressor beams, a major justification for undertaking this exercise is to push general relativity to the breaking point; in the hope that the resulting wreckage will tell us something interesting --- possibly even about quantum gravity~\cite{Book,Lobo:2017}.

\enlargethispage{40pt}
After first analysing Natário's  generic warp drive case~\cite{Natario:2001}, we will consider three special cases:
\begin{enumerate}[nosep]
	\item 
	We modify the Alcubierre fixed-flow-direction warp field.
	\item
	We modify the Natário zero-expansion warp field.
	\item
	We modify the zero-vorticity warp field.
\end{enumerate} 
We shall also illustrate each of these three cases with some specific examples based on beams with a Gaussian profile.

A recurring theme in the analysis will be the use of the classical point-wise energy conditions (null, weak, strong, and dominant; abbreviated NEC, WEC, SEC, and DEC, respectively)~\cite{Hawking-Ellis,Curiel:2014,Kontou,Martin-Moruno:2013-quantum1}. They can be considered as an attempt to remain as agnostic as possible about underlying equations of state. While the energy conditions do not seem to be fundamental physics, they are at the very least a very good sanity check on just how weird the physics is getting~\cite{Curiel:2014,twilight,trace}. We already know of examples of violations at microscopic scales (\emph{e.g.,} Hawking radiation) and mesoscopic scales (\emph{e.g.,} Casimir effect). No macroscopic violations of the energy conditions are known up to this point, except at truly cosmological scales -- and they violate only \emph{some} of the energy conditions (the accelerated expansion of the universe violates the strong and dominant energy conditions, but not the null and weak energy conditions~\cite{cosmo99,epoch,science,galaxy}).  
Therefore, besides the violation of the energy conditions not being an absolute prohibition, it is an indication that one should look very carefully at the underlying physics~\cite{twilight,trace}.
For more background on the energy conditions see~\cite{FEC1,FEC2,
Martin-Moruno:2015-quantum2, gvp1,gvp2,gvp3,gvp4,gvp5,Fewster:2002,Fewster:2010,Fewster:2012,
scale-anomalies,Flanagan:1996,Ford:1994,Ford:1995,Ford:2003,Ford:2005,
Wald:anec,Visser:2003,Hochberg:1998a,Hochberg:1998b,Barcelo:2000, Barcelo:brane, Kar:2004,
Roman:1983, Roman:wormhole, Lobo:2016, Hochberg:1998c}.

For the sake of full transparency, we should also mention that our interest in these topics was rekindled and inspired by three recent papers~\cite{Lentz:2020, Bobrick:2021, Fell:2021}. Unfortunately, significant parts of those three papers are incorrect, misguided, and/or  misleading. See reference~\cite{Santiago:2021} for details. 

\addtocontents{toc}{\protect\vspace{-5pt}}
\section{When things need to be moved}

\begin{figure}
	\centering
	\begin{subfigure}[t]{\textwidth}
		\includegraphics[width=\textwidth]{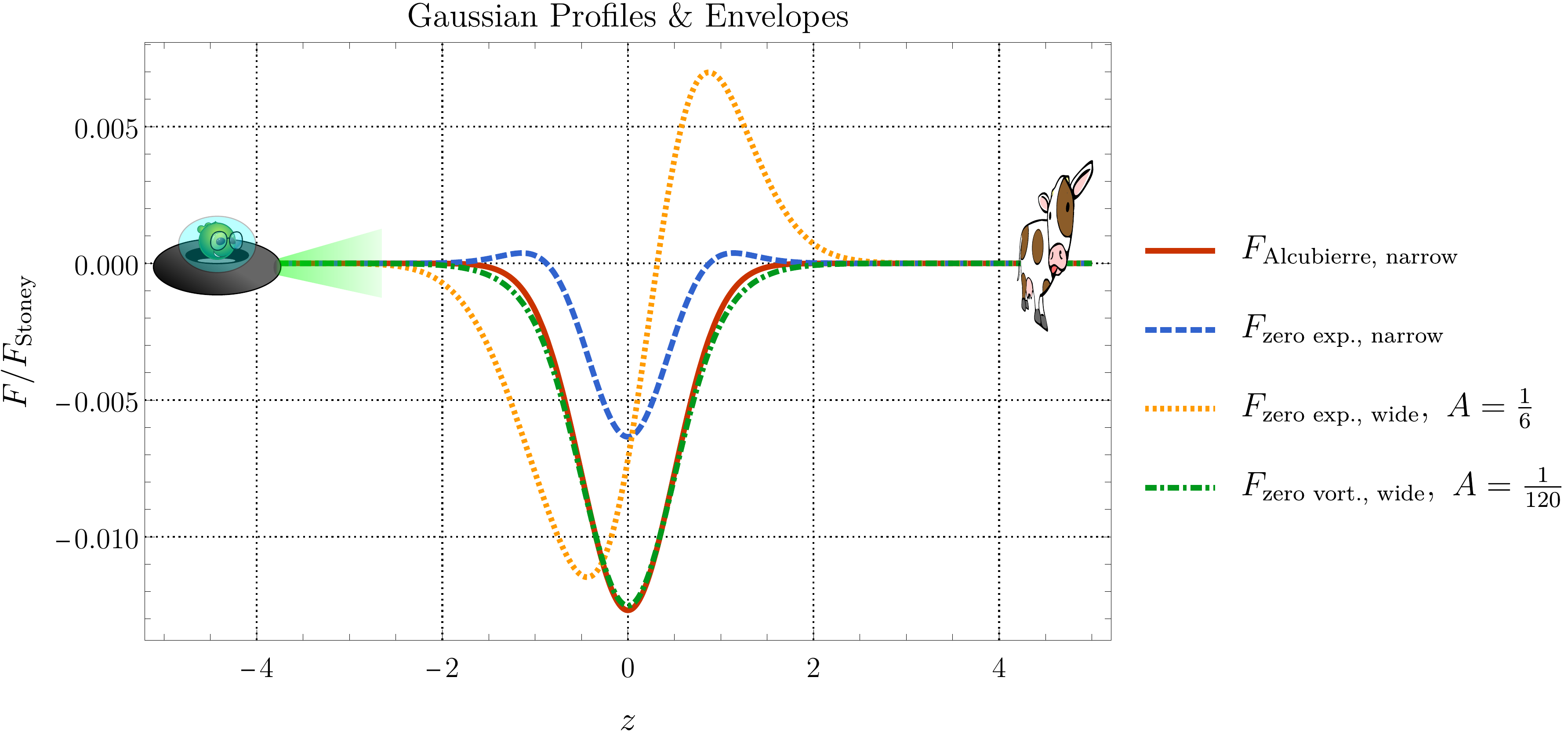}
		\caption{}
		\label{fig:GaussF}    
	\end{subfigure}\\
	\begin{subfigure}[t]{\textwidth}
		\includegraphics[width=.33\textwidth]{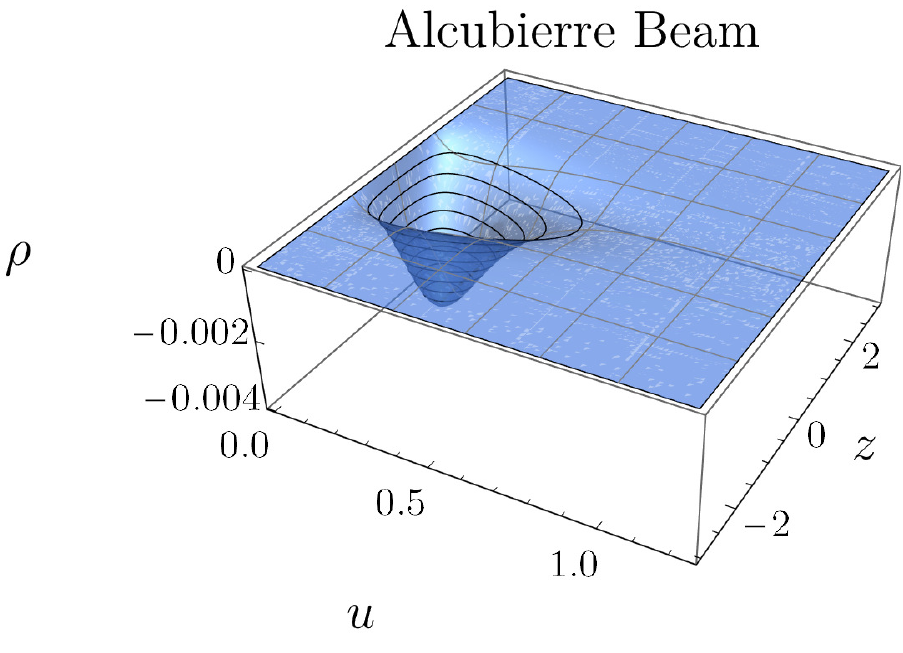}~
		\includegraphics[width=.33\textwidth]{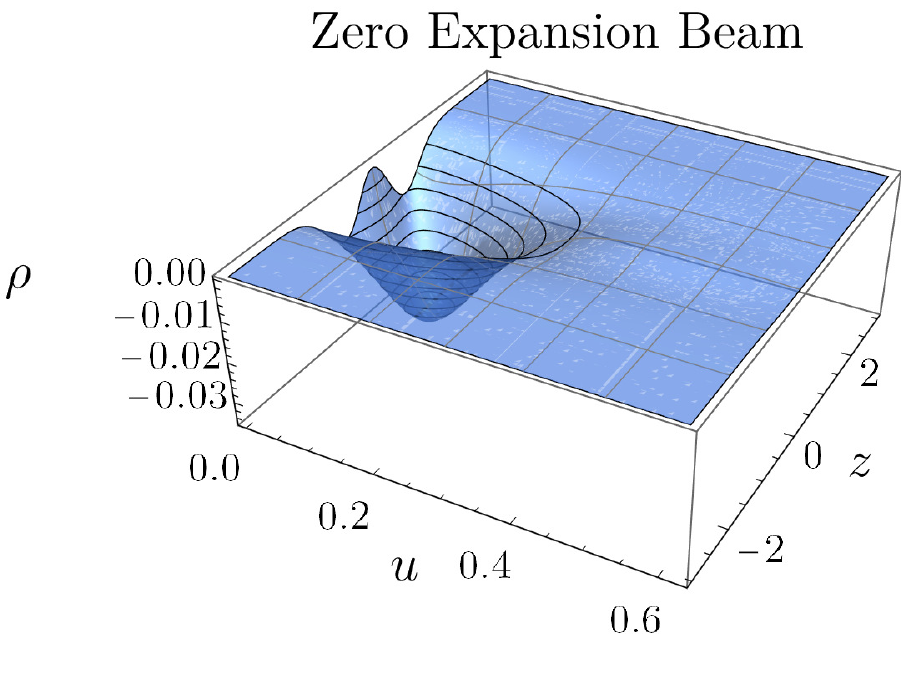}~
		\includegraphics[width=.33\textwidth]{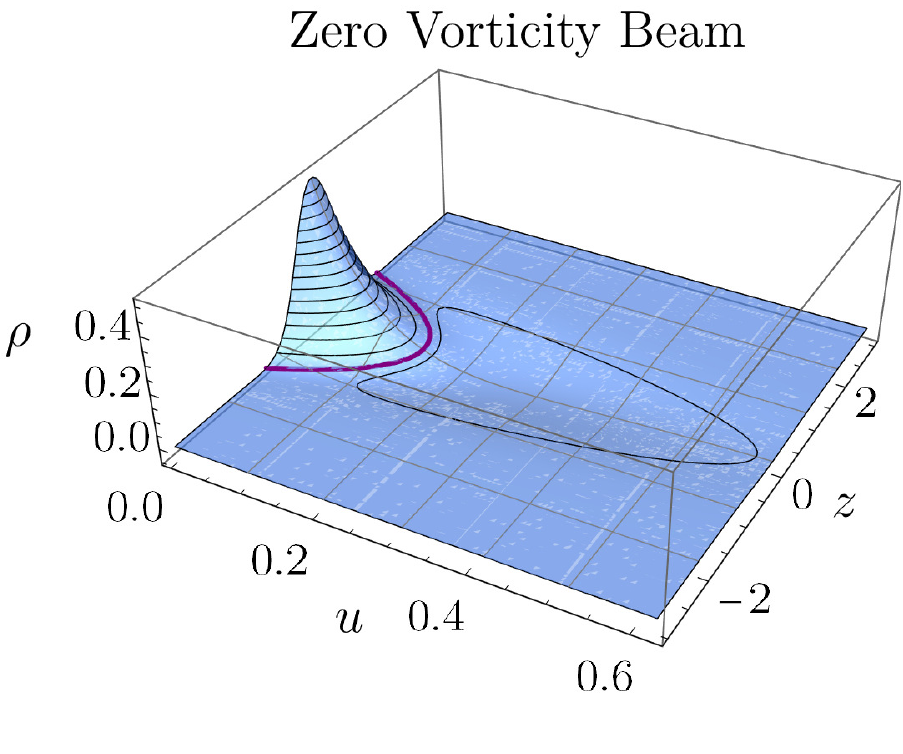}~
		\caption{}
		\label{fig:GaussRho}    
	\end{subfigure}	\caption{{Forces (above) and energy densities (below) for Gaussian beam profiles. The field is assumed to be sourced by someone on the left at negative $z$, the target---a flat cow in the tractor field space-time---on the right at positive $z$. Choosing the source and target provides for a distinction between \emph{tractor} and \emph{pressor} (or \emph{repulsor}) fields. Details concerning this particular beam configuration can be found in section~\ref{sec:GaussEnv}. The parameters of equation~\eqref{E:GaussPlot} that we have chosen are: $A=0.5$, $B=C=1.0.$ The purple line in the density plot for the zero-vorticity beam indicates the location where the energy density is zero.}}
	\label{fig:Gauss}
\end{figure}

One of Wheeler's adages that became standard general relativity folklore is the famous saying that \enquote{space-time tells matter how to move; matter tells space-time how to curve}. From many a practical point of view, questions regarding objects' movement are less about the \emph{how} and more about the \emph{ought} --- things are wanted elsewhere from where they are now. It is this logistical perspective that we shall address in the following: How can we ensure that general relativity does the job of moving an object (like a cow, [preferably a spherical cow, in vacuum],  or a Corellian CR90 corvette) for us? 

The key ingredient will be to limit ourselves to test field cases, where we neglect the mass of the objects we want to move, how they interact with space-time and with the matter we put in space-time to move them. This reduces the core physics question to one of forces: We want to use the pressures encoded in the stress-energy tensor of a beam-like field to move target test masses.

The primary force-related calculation we shall undertake is this:  If the beam is pointed in the $z$ direction, then one calculates the stress-energy component $T_{zz}(t,x,y,z)$,
and integrates it over the entire transverse $x$-$y$ plane to find the net force:
\begin{equation}
\label{Fztotal}
F(t,z) = \pm \int_{\mathbb{R}^2} T_{zz}(t,x,y,z) \, \d x \, \d y.
\end{equation}
Here the $+$ sign corresponds to a beam impinging on the target from the left, whereas the $-$ sign corresponds to a beam impinging on the target from the right.
There is an approximation being made here, that the beam is \emph{narrow} with respect to the target, so that it is a good approximation to integrate over the \emph{entire} transverse $x$-$y$ plane. If the beam is instead \emph{wide} compared to the size of the target then one should instead use the approximation
\begin{equation}
\label{FzA}
F(t,z) = \pm  T_{zz}(t,0,0,z) \, A.
\end{equation}
Here $T_{zz}(t,0,0,z)$ is the on-axis stress, and $A$ is the cross sectional area of the target.
For a beam of intermediate widths, (comparable to the size of the target), one would in principle need to calculate
\begin{equation}
	\label{FzintA}
	F(t,z) = \pm  \int_A  T_{zz}(t,x,y, z) \, \d x \; \d y,
\end{equation}
but this is unnecessarily complicated for the primary issues we wish to address. 
\enlargethispage{20pt}

The quantity $F(t,z)$ is the \emph{net} force the beam exerts on some target located at some position $z$ at time $t$. For convenience we shall henceforth assume that the field is generated by someone positioned on the left, and that the target 
be positioned to the right of the generator, (see Figure \ref{fig:Gauss}), thus allowing us to restrict attention to the plus sign in equations (\ref{Fztotal}) -- (\ref{FzintA}). We shall furthermore assume that the target will move under the influence of the field, while the \enquote{generator} will not, and---as mentioned above---both behave as test fields. This setup provides for a simple characterization of the effect of the field:
If $F(t,z)<0$, corresponding to attraction, we call this a \emph{tractor beam}.
If $F(t,z)>0$, corresponding to repulsion, we call this a \emph{pressor beam}\footnote{{Again, this definition depends on the relative positions of generator and target, we assume the target is always to the right of the generator.}}.
On the other hand, the definition of a \emph{stressor beam} can be a little trickier. The reason for this being that, independent of the overall sign of $F(t,z)$, one can quite generally define a beam which has significantly varying pressure across the cross-sectional area of the target. In this way, there might be a certain ambiguity about when a specific beam would be considered to be a tractor/pressor or a stressor beam, since this would depend on the properties of the target material --- such as its elasticity and ultimate yield strength, and so on. 

However, for most ``applications'', we expect the $T_{zz}$ component for a tractor (pressor) beam to not vary too greatly over its region of influence on the target. 
A quick measure of when a beam would behave as a stressor beam is given by:
\begin{equation}
\max\nolimits_A (T_{zz}) - \min\nolimits_A (T_{zz}) \gg \sigma_{\text{material}}.
\end{equation}
Here $A$ is the cross-sectional area of the target exposed to the beam, 
and  $\sigma_{\text{material}}$ is the ultimate yield stress of the material making up the target.

While equations \eqref{Fztotal}--\eqref{FzintA} are universally valid, both for standard general relativity, and for modified theories of gravity, we will focus mainly on standard general relativity. Therefore, using the Einstein equations, we have for a narrow beam  in terms of the Einstein tensor:
\begin{equation}
	F(t,z) = {1\over 8\pi} \int_{\mathbb{R}^2} G_{zz}(t,x,y,z) \, \d x \, \d y,
\end{equation}
while for a wide beam
\begin{equation}
	F(t,z) = {1\over 8\pi} \; G_{zz}(t,0,0, z) \, A.
\end{equation}
These are the key equations we will be using in the following sections.

As usual, we are using geometrodynamic units, where $G_\mathrm{Newton}\to 1$ and $c\to 1$. If one wishes to reinstate SI units, then in terms of the Stoney force $F_*=c^4/G_\mathrm{Newton}$~\cite{Stoney1,Stoney2,Stoney3,Jowsey:2021a,Jowsey:2021b}~\footnote{The Stoney force  happens to equal the Planck force $F_* = E_\mathrm{Planck}/L_\mathrm{Planck}$, (the various factors of $\hbar$ cancel~\cite{Jowsey:2021a,Jowsey:2021b}).} one has
\begin{equation}
	F(t,z) = {F_*\over 8\pi} \int_{\mathbb{R}^2} G_{zz}(t,x,y,z) \, \d x \, \d y,
\end{equation}
and
\begin{equation}
	F(t,z) = {F_*\over 8\pi} \; G_{zz}(t,0,0, z) \, A.
\end{equation}

It is worthwhile mentioning that the magnitude of the Stoney force is truly enormous --- some $1.2 \times 10^{44}$ Newtons.  Accordingly, relatively small spacetime curvatures (weak-field gravity) can still lead to significant human-scale forces and stresses. It is beyond the scope of the present article to consider just how weak the weak fields can be before the test field approximations for the target mass break down.

\clearpage
\addtocontents{toc}{\protect\vspace{-5pt}}
\section{Natário's generic warp field}\label{sec: Natario}
\subsection{Kinematics}

\enlargethispage{20pt}
Our tractor/pressor/stressor beams will be based on modifications of Natário style \emph{generic} warp drives~\cite{Natario:2001,Lobo:2004,Lobo:2017, Alcubierre:2017,Santiago:2021}. The generic form of the space-time metric line element is 
\begin{equation} 
\label{E:line}
     \d s^2 = - \d t^2 + \delta_{ij} \; \left( \d x^i - v^i(t,x,y,z)\, \d t\right) \; \left( \d x^j - v^j(t,x,y,z) \,\d t\right).  
\end{equation}
Note that the lapse is unity, $N\to 1$, the spatial slices are flat, $g_{ij} \to \delta_{ij}$, 
and the ``flow'' vector $v^i(t,x,y,z)$ is the negative of what is (in the ADM decomposition) usually called the ``shift'' vector~\cite{ADM2,Poisson,Alcubierre:3+1,ADM1}. 

A kinematically useful quantity is the vorticity of the flow field, 
\begin{equation}\label{E:vorticity}
\vec\omega = \curl \vec v
\end{equation}
and its square, $\vec \omega\cdot\vec \omega$.

The constant-$t$ spatial slices have covariant normal $n_a=\partial_a t  = (-1,0,0,0)_a$, whose contravariant components are the future-pointing 4-velocity $n^a=(1,v^i)$. Observers that ``go with the flow'', moving with 4-velocity $n^a$, are geodesics, and are often called Eulerian.

The intrinsic geometry of the spatial slices is flat, whereas the extrinsic geometry is characterized  by the second fundamental form $K_{ij} = v_{(i,j)}$. The components of the Riemann tensor can then be evaluated in terms of the extrinsic curvature and the normal. For details see, for instance,~\cite{Santiago:2021, ADM2, Poisson, Alcubierre:3+1,ADM1}. For \emph{even more} additional background on ``warp field'' space-times see references~\cite{Everett:1995, Everett:1997, Hiscock:1997, Pfenning:1998, Low:1998, Olum:1998, VanDenBroeck:1999a, VanDenBroeck:1999b, Clark:1999,Visser:1999, Alcubierre:2001, Lobo:2002, Lobo:2004b, Lobo:2007, Finazzi:2009, McMonigal:2012}.

\subsection{Stress-energy tensor}

 For current purposes, adopting the unit-lapse and flat-spatial-slices conditions,   key results for the components of the stress-energy tensor are ~\cite{Alcubierre:3+1,Santiago:2021,ADM2,Poisson}:
\begin{itemize}
\item  The Gauss--Codazzi equations yield the Eulerian energy density:
\begin{equation}
\label{E:rho}
\rho = {G_{nn}\over 8\pi} = {G_{ab} n^a n^b \over 8\pi} = {1\over16\pi} \left(  K^2-\tr(K^2) \right),
\end{equation}
where $K = K_{ij}\delta^{ij} = \tr(K)$ and $\tr(K^2) = (K^2)_{ij}\delta^{ij} = K_{ik}\delta^{kl}K_{lj}\delta^{ij}$. 

In the current context this can be recast as~\cite{Santiago:2021}:
\begin{equation}
\label{E:divergence-x}
\rho =
{1\over16\pi} \left\{ \div \{ \vec v\,K - (\vec v \cdot\grad)  \vec v\}  - {1\over2} \, (\vec\omega \cdot \vec\omega) \right\}.
\end{equation}

\item The Gauss--Mainardi equations yield the Eulerian energy flux:
\begin{equation}
\label{E:flux}
f_i = {G_{ni}\over 8\pi}=  {G_{ai} \,n^a\over 8\pi}={1\over 16\pi} \left( \curl (\curl \vec v) \right)_i.
\end{equation}
\item 
The $3\times3$ stress tensor is somewhat messier, and can be expressed in terms of the extrinsic curvature and its Lie derivatives~\cite{Santiago:2021}: 

\begin{equation}
 \label{E:stress}
T_{ij} = {G_{ij}\over 8\pi} =
 {1\over8\pi} \left( \L_n K_{ij}  + K K_{ij} - 2(K^2)_{ij} -  \left(\L_n K  + {1\over2}K^2 + {1\over2} \tr(K^2)\right) \delta_{ij}\right).
\end{equation}
For the various explicit examples we consider below, we shall instead often use \emph{ab~initio} calculations instead of this general (but relatively intractable) result. 
\item
In contrast, the trace of the $3\times3$ stress tensor is somewhat easier to deal with. For the average pressure $\bar p$ we have~\cite{Santiago:2021}:
\begin{equation}
\label{E:bar-p}
\bar p = {T_{ij}\; \delta^{ij} \over 3} =  -{1\over24\pi} \left(  2 \nabla_a ( K n^a) -{3\over2} K^2 + {3\over2} \tr(K^2) \right). 
\end{equation}
This implies
\begin{equation}
\label{E:bar-p2}
\bar p = \rho   -{1\over12\pi}\nabla_a ( K n^a). 
\end{equation}
\end{itemize}

These are the key stress-energy components we need for the current task. 
For further discussion on these and related issues see references~\cite{Santiago:2021,ADM2, Poisson, Alcubierre:3+1}.  

An immediate consequence of these general results is that once appropriate fall-off conditions are imposed at spatial infinity one has
\begin{align}
\int \rho\;\d^3 x &= - {1\over32\pi} \int (\vec\omega \cdot \vec\omega) \;\d^3 x \leq 0,\label{E:intrho}\\
\int (\rho +p) \d^3 x &= - {1\over16\pi} \int (\vec\omega \cdot \vec\omega) \;\d^3 x \leq 0. 
\end{align}
This implies that violations of the WEC and NEC are unavoidable~\cite{Santiago:2021}, 
and we shall see similar results repeatedly recurring in the subsequent discussion.

\addtocontents{toc}{\protect\vspace{-5pt}}
\section{Beam profile}\label{sec:Beam profile}

In this section, we will discuss the kinematics and general properties of the stress-energy tensor of such beams, including the forces key to our interpretation of them.

\subsection{Beam kinematics}

For our purposes we shall choose a factorized ``beam'' profile for the flow vector, one that respects axial symmetry around the $z$-axis:
\begin{align}
v_x(t,x,y,z) &=  k(t,z)\, x\, h(x^2+y^2),
\label{E:beam1}\\
v_y(t,x,y,z) &=   k(t,z)\, y\, h(x^2+y^2),
\label{E:beam2}\\
v_z(t,x,y,z) &= v(t,z)\, f(x^2+y^2).
\label{E:beam3}
\end{align}
We shall refer to $f(x^2+y^2)$ and $h(x^2+y^2)$ as profile functions, whereas $v(t,z)$ and $k(t,z)$ will be referred to as envelope functions. 

Note the explicit presence of $x$ and $y$ in the flow components $v_x$ and $v_y$,  precisely to maintain axial symmetry.
Furthermore, 
\begin{equation}
\sqrt{v_x(t,x,y,z)^2 + v_y(t,x,y,z)^2} = k(t,z)\, \sqrt{x^2+y^2} \, h(x^2+y^2). 
\end{equation}
Useful definitions of the average transverse width of the beam are to consider
\begin{equation}
W_{xy}^2 = {\int (x^2+y^2) \,(v_x^2+v_y^2) \,\d x \d y \over \int \,(v_x^2+v_y^2) \,\d x \d y }
= {\int (x^2+y^2)^2 \,h(x^2+y^2)^2 \, \d x \d y \over  \int (x^2+y^2) \,h(x^2+y^2)^2 \, \d x \d y },
\end{equation}
and/or
\begin{equation}
W_z^2 = {\int (x^2+y^2) \, (v_z^2)\, \d x \d y \over \int (v_z^2) \,\d x \d y }
= {\int (x^2+y^2) \,f(x^2+y^2)^2\, \d x \d y \over  \int  f(x^2+y^2)^2\, \d x \d y }.
\end{equation}
Both of these characterizations of average width depend only on the profile functions, not on the envelope functions.
\enlargethispage{10pt}

Far away from the beam axis, as $x^2+y^2\to\infty$, we will demand that both profile functions tend to zero: $f(x^2+y^2)\to 0$ and $h(x^2+y^2) \to 0$, in order that the beam asymptotically reduces to flat Minkowski space.  
All of the $t$ and $z$ dependence is  encoded in the two functions $v(t,z)$ and $k(t,z)$.
Since one wants the beam to be of finite length, and not stretch all the way across the universe, one should demand both $\lim\limits_{z\to\pm\infty}v(t, z) \to 0$ and $\lim\limits_{z\to\pm\infty}k(t, z) \to 0$, again ensuring an asymptotic approach to Minkowski space. 

More precisely, we shall demand sufficiently rapid fall-off at spatial infinity, which will then also allow integration by parts unrestricted by boundary terms. We shall also enforce smooth on-axis behaviour by demanding that the profile functions and their derivatives be finite on the beam axis. These structural assumptions for the flow vector is basically our definition of what we mean by a ``beam'' directed along the $z$-axis. 

The previously introduced vorticity~\eqref{E:vorticity} for our beam geometry reduces to:
\begin{alignat}{3}
	\omega_x &=& \partial_y v_z - \partial_z v_y 
		&=& \; -& y \left\{ \partial_z k(t,z) h(x^2+y^2) - 2 v(t,z) f'(x^2+y^2) \right\},\\
	\omega_y &=& \;-\partial_x v_z + \partial_y v_y 
		&=&  &x \left\{ \partial_z k(t,z) h(x^2+y^2) - 2 v(t,z) f'(x^2+y^2) \right\},\\
	\omega_z &=& \partial_x v_y - \partial_y v_x &=& &0. 
\end{alignat}
The square of the vorticity,
\begin{equation}
	\label{E:omega-sq}
	\vec \omega\cdot\vec \omega 
	= (x^2+y^2) \Big\{ \partial_z k(t,z) h(x^2+y^2) - 2 v(t,z) f'(x^2+y^2) \Big\}^2,
\end{equation}
will show up quite often in subsequent calculations.

\enlargethispage{10pt}

\subsection{Stress-energy basics}
\enlargethispage{10pt}
If we now additionally impose the factorization conditions \eqref{E:beam1}--\eqref{E:beam2}--\eqref{E:beam3} appropriate to a beam geometry,  then the axial symmetry imposes additional constraints on the stress-energy tensor. Specifically:
 \begin{equation}
{T_{xx}-T_{yy}\over T_{xy}} = {x^2-y^2\over xy},
\end{equation}
and
\begin{equation}
{T_{xz}\over T_{yz}} = {T_{nx}\over T_{ny}}= {x\over y}.
\end{equation}
This implies in particular that 
\begin{equation}\label{E:Txy}
\left[\begin{array}{cc} T_{xx} & T_{xy} \\ T_{xy} & T_{yy} \end{array} \right]
=
\left[\begin{array}{cc} x^2 & xy \\ xy & y^2 \end{array} \right]  \mathcal{F}_1(t,x,y,z) 
+
\left[\begin{array}{cc} 1 &0 \\ 0 & 1 \end{array} \right]  \mathcal{F}_2(t,x,y,z), 
\end{equation}
and
\begin{equation}\label{E:Tiz}
\left[\begin{array}{cc} T_{xz} & T_{yz}  \end{array} \right]
=
\left[\begin{array}{cc} x & y \end{array} \right]  \mathcal{F}_3(t,x,y,z).
\end{equation}
Similarly, for the $x$-directed and $y$-directed fluxes, we have:
\begin{equation}
\left[\begin{array}{cc} f_{x} & f_{y} \end{array} \right]
=
\left[\begin{array}{cc} T_{nx} & T_{ny} \end{array} \right]
=
\left[\begin{array}{cc} x & y  \end{array} \right]  \mathcal{F}_4(t,z,y,z).
\end{equation} 
The $\mathcal{F}_i(t,x,y,z)$ are specific scalar functions that can be explicitly calculated when required.
However, the $\mathcal{F}_i(t,x,y,z)$ are not the most interesting quantities for our purposes. We shall instead be more focussed on the comoving energy density $\rho(t,x,y,z)$,  the stress-energy component $T_{zz}(t,x,y,z)$, the flux component $f_z(t,x,y,z)$ directed along the beam axis,
and the average stress $\bar p(t,x,y,z)$.

We now continue our calculations using the generic beam-like flow \eqref{E:beam1}--\eqref{E:beam2}--\eqref{E:beam3}. As yet, we impose no extra restriction on the four functions $v(t,z)$, $k(t,z)$, $f(x^2+y^2)$, and $h(x^2+y^2)$, apart from the previously mentioned asymptotic conditions. Namely that $f(x^2+y^2)\to 0$ and $h(x^2+y^2) \to 0$ away from the beam axis and both $\smash{\lim\limits_{z\to\pm\infty}} v(t, z) \to 0$ and $\lim\limits_{z\to\pm\infty} k(t, z) \to 0$.

\subsection{Force}
In order to calculate the force~\eqref{Fztotal}, let us now investigate $T_{zz}(t,x,y,z)$ for this factorized flow, and integrate this over the $x$-$y$ plane. For $T_{zz}(t,x,y,z)$ we find: 
\begin{align}
\label{E:generic-Tzz}
	T_{zz}(t,x,y,z) &= {1\over8\pi} \left\{ N_1(x,y) \;v(t,z)^2 + N_2(x,y)  \;v(t,z) \partial_z k(t,z)+ 
	N_3(x,y) \;[\partial_z k(t,z)]^2
	\right.
	\nonumber\\
	& \qquad \qquad \left.
	+ N_4(x,y) \;\partial_t k(t,z)+ N_5(x,y) \; k(t,z)^2 \right\}.
\end{align}

Here, using the shorthand $u= x^2+u^2$, we have:
\begin{align}
	N_1(x,y) &= - 3 u [f'(u)]^2,\label{E:generic-N1}\\
	N_2(x,y) &= 
	- 2 [u f(u) h(u)]' + u h(u) f'(u),\label{E:generic-N2}\\ 
	N_3(x,y) &= \hphantom{-}{1\over4} u [h(u)]^2,\label{E:generic-N3}\\
	N_4(x,y) &= -2 [u h(u)]',\label{E:generic-N4}\\
	N_5(x,y) &= -3 [u h(u)^2]'- 4 [u^2 h(u) h'(u)]' .\label{E:generic-N5}
\end{align}
Without detailed calculation we can immediately deduce:
\begin{equation}
	\int_\R2 N_1(x,y)  \, \d x \d y<0, \qquad \int_\R2 N_2(x,y)  \, \d x \d y= \hbox{indefinite}, 
	\quad 
\end{equation}
\begin{equation}
        \int_\R2 N_3(x,y)  \, \d x \d y>0, \qquad 
	\int_\R2 N_4(x,y)  \, \d x \d y= \int_\R2 N_5(x,y)  \, \d x \d y= 0. 
\end{equation}

It is worth noting that
\begin{equation}
		\int_\R2 N_1(x,y)  \, \d x \d y= -3 \pi\int_0^\infty u [f'(u)]^2\d u , 
	\end{equation}
\begin{equation}
		\int_\R2 N_2(x,y)  \, \d x \d y= \pi\int_0^\infty u h(u) f'(u)\d u  ,
	\end{equation}
		and
	\begin{equation}
		\int_\R2 N_3(x,y)  \, \d x \d y= \pi \int_0^\infty {1\over4} u [h(u)]^2\d u. 
	\end{equation}
	Using this, we find that in the narrow beam approximation
	\begin{align}
		\label{E:critical} 
		F(t,z) &= 
		\int_\R2 T_{zz}(t,x,y,z)  \; \d x \d y\\
		&= -{1\over2} v(t,z)^2 \int_0^\infty u [f'(u)]^2\d u + {1\over 8} 
		\int_0^\infty u \left[ v(t,z) f'(u) + {1\over2} \partial_z k(t,z) h(u)\right]^2\d u.
	\end{align}	
This is a sum of negative definite and positive definite terms, thus allowing the generic beam to potentially be fine-tuned as either a tractor or a pressor (or even a stressor). 

In contrast, in the wide beam approximation we need to evaluate $T_{zz}(t,0,0,z)$. Note 
\begin{gather}
N_1(0,0)=0, \quad N_2(0,0)=-2f(0)h(0), \quad N_3(0,0)=0, \nonumber\\
N_4(0,0)=-2h(0), \qquad N_5(0,0)=-3h(0)^2. 
\end{gather}
Consequently 
\begin{equation}
\label{E:generic-Tzz0}
	T_{zz}(t,0,0,z) = -{h(0)\over8\pi} \left\{2f(0) \;v(t,z) \partial_z k(t,z)+ 
	2\;\partial_t k(t,z)+3h(0) \; k(t,z)^2 \right\}.\qquad
\end{equation}
So in the wide-beam approximation the force exerted on the target is 
\begin{equation}
\label{E:generic-wide}
        F(t,z) = -{h(0)\over8\pi} \left\{2f(0) \;v(t,z) \partial_z k(t,z)+ 
	2\;\partial_t k(t,z)+3h(0) \; k(t,z)^2 \right\} A.\qquad
\end{equation}
This is of indefinite sign, depending delicately on the envelope functions, potentially allowing either tractor/pressor behaviour.

\subsection{Flux}

The flux in the $z$-direction, as defined in equation~\eqref{E:flux}, is given by:
\begin{equation}
	f_z(t,x,y,z) = {1\over8\pi} \left\{- \partial_z k(t,z) [u h(u)]' + 2 v(t,z) [u f'(u)]'\right\}.
\end{equation}
Thence,
\begin{align}
	\int_\R2 f_z(t,x,y,z) \d x \d y &= {1\over8} \int_0^\infty 
	\left\{ -\partial_z k(t,z) [u h(u)]'  + 2 v(t,z) [u f'(u)]'\right\}\d u = 0.
\end{align}

For the $x$-direction
\begin{align}
	f_x(t,x,y,z) &= -{x\over8\pi} \left\{\partial_z v(t,z)  f'(x^2+y^2)  -{1\over2}\partial_z^2 k(t,z)  h(x^2+y^2)\right\}.
\end{align}
Anti-symmetry under $x \longleftrightarrow -x$ now yields
\begin{equation}
	\int_\R2 f_x(t,x,y,z) \d x \d y= 0.
\end{equation}
Similarly, for the $y$-direction
\begin{equation}
	f_y(t,x,y,z) = -{y\over8\pi} \left\{\partial_z v(t,z)  f'(x^2+y^2)  -{1\over2}\partial_z^2 k(t,z) yh(x^2+y^2)\right\},
\end{equation}
and again by appealing to anti-symmetry,
\begin{equation}
	\int_\R2 f_y(t,x,y,z) \d x \d y= 0.
\end{equation}
\enlargethispage{30pt}

Consequently, for the general tractor/pressor/stressor beam we \emph{always} have the net flux integrating to zero:
\begin{equation}
	\int_\R2 f_i(t,x,y,z) \d x \d y= 0.
\end{equation}
Thence, at least in the narrow-beam approximation, we never need to worry about the net fluxes impinging on the target, they always quietly cancel. 
However, even if the net fluxes seen by Eulerian observers cancel, there might be significant fluctuations around zero over the cross-sectional area of the target. 
For instance, on axis we have
\begin{equation}
\vec f(t,0,0,z) =  {1\over8\pi} \left\{- \partial_z k(t,z) h(0)+ 2 v(t,z) f'(0)\right\} \hat z.
\end{equation}
It is now the envelope functions $v(t,z)$ and $k(t,z)$ that primarily drive the localized on-axis fluxes in the wide-beam approximation.

\subsection{Off-diagonal stress components}

Similar steps can be applied to equation~\eqref{E:Tiz} concerning the $T_{xz}$ and $T_{yz}$ components:
\begin{equation}
	T_{xz} (t,x,y,z) = x \F_3(t,x,y,z), \qquad T_{yz} (t,x,y,z) = y \F_3(t,x,y,z),
\end{equation}
implying (using anti-symmetry under $x \longleftrightarrow -x$ and $y \longleftrightarrow -y$ respectively)
\begin{equation}
	\int_\R2 T_{xz} (t,x,y,z) \d x \d y= \int_\R2 T_{yz} (t,x,y,z) \d x \d y=0.
\end{equation}

Finally, from equation~\eqref{E:Txy} we get:
\begin{equation}
	T_{xy} (t,x,y,z) = xy\,  \F_1(t,x,y,z), 
\end{equation}
implying (now using \emph{either} anti-symmetry under $x \longleftrightarrow -x$, \emph{or} anti-symmetry under $y \longleftrightarrow -y$)
\begin{equation}
	\int_\R2 T_{xy} (t,x,y,z) \d x \d y=0.
\end{equation}

Combining all the above, the integral $\int_\R2 T_{\hat a\hat b} \, \d x \d y$ is purely diagonal, all off-diagonal elements vanish:
\begin{equation}\label{E:intTmatrix}
	\int_\R2 T_{\hat a\hat b}\; \d x \d y= \left[\begin{array}{cccc}
		\int_\R2 \rho \; \d x \d y&0&0&0\\
		0& \int_\R2 T_{xx} \; \d x \d y&0&0\\
		0&0 & \int_\R2 T_{xx} \; \d x \d y&0\\
		0&0 & 0 &\int_\R2 T_{zz} \; \d x \d y
	\end{array}\right]. 
\end{equation}
This really is just a consequence of the assumed axial symmetry of our beam.
These observations have the effect of focussing our attention on the diagonal components of the (integrated) stress-energy.

\subsection{Eulerian energy density}
\enlargethispage{20pt}
For the Eulerian comoving energy density in this generic beam we find:
	\begin{align}
		\rho(t,x,y,z) &= {1\over8\pi} \left\{ R_1(x,y)  v(t,z)^2 
		+ R_2(x,y) v(t,z) \partial_z k(t,z)   + R_3(x,y) k(t,z) \partial_z v(t,z) 
		\right.\nonumber\\
		&\left. \qquad \qquad
		+ R_4(x,y) [\partial_z k(t,z)]^2   + R_5(x,y)  k(t,z)^2
		\right\},
	\end{align}
	with
	\begin{align}
		R_1(x,y)  &= - u [f'(u)]^2,\\
		R_2(x,y)  &= -u h(u) f'(u),\\
		R_3(x,y)  &= \hphantom{-}2 f(u) [u h(u)]',\\
		R_4(x,y)  &= - u h(u)^2/4,\\
		R_5(x,y)  &= \hphantom{-}[u h(u)^2]' .
	\end{align}	
	Then, after an integration by parts,
	\begin{equation}
		\int_\R2 R_3(x,y) \d x \d y=  2 \int_\R2 R_2(x,y) \d x \d y= - 2 \pi \int_0^\infty  u h(u) f'(u).
	\end{equation}
	Furthermore, 
	\begin{equation}
		\int_\R2 R_5(x,y) \d x \d y= 0.
	\end{equation}
	Thence,
	\begin{align}
		\int_\R2 \rho(t,x,y,z) \d x \d y =& {1\over 8} \left\{ \partial_z [v(t,z) k(t,z)] \int_0^\infty  u h(u) f'(u)\right.
		\nonumber\\
		&
		\left.
		- \left[ v(t,z)^2  \int_0^\infty u [f'(u)]^2\d u + k(t,z) \partial_z v(t,z) \int_0^\infty u f'(u) h(u)\d u
		\right.\right.
		\nonumber\\
		&\left.\left. 
		+ {1\over4} [\partial_z k(t,z)^2] \int_0^\infty u h(u)^2\d u
		\right] 
		\right\} .
\end{align}
Now we also integrate over $z$ and apply appropriate boundary conditions at $z=\pm \infty$ (where the beam has to switch off by definition) to discard the first term, which is a total derivative. Then,
\begin{align}
		\int_\R3 \rho(t,x,y,z) \d x \d y \d z =& 
		-{1\over 8}\int_{-\infty}^{+\infty}  \left\{  v(t,z)^2  \int_0^\infty u [f'(u)]^2\d u
		\right.
		\nonumber\\
		&\qquad\qquad
		 +k(t,z) \partial_z v(t,z) \int_0^\infty u f'(u) h(u)\d u
		\nonumber\\
		&\qquad\qquad
		 \left.
		+ {1\over4} [\partial_z k(t,z)^2] \int_0^\infty u h(u)^2\d u
		\right\}  \d z.
\end{align}

\subsection{Weak energy condition}

This puts us into a good position to have a first look at an energy condition, this time the WEC. Let us do another integration by parts, again invoking suitable boundary conditions,  to replace $\int_{-\infty}^{+\infty}  k(t,z) \partial_z v(t,z) \d z \to 
- \int_{-\infty}^{+\infty}  v(t,z) \partial_z k(t,z)\d z$. 

But this is now actually a perfect square:
	\begin{equation}
		\label{E:square}
		\int_\R3 \rho(t,x,y,z) \d x \d y \d z = -{1\over 8} \int_0^\infty  \int_{-\infty}^{+\infty} 
		u \left[f'(u)  v(t,z) -{1\over2} h(u) \partial_z k(t,z) \right]^2 \d z\d u \leq 0.
	\end{equation}
The integrand appearing above is just ${1\over4}$ of the square of the vorticity $(\vec\omega\cdot\vec\omega)$, see equation~\eqref{E:omega-sq}, so that this is equivalent to	
\begin{equation}
	\label{E:square2}
	\int_\R3 \rho(t,x,y,z) \;\d x \d y \d z = -{1\over 32} \int_0^\infty  \int_{-\infty}^{+\infty} 
	(\vec\omega\cdot\vec\omega)\; \d z\d u \leq 0.
\end{equation}	
This should not come as a surprise, given it is just equation~\eqref{E:intrho}.

Accordingly, in this generic tractor/pressor/stressor beam configuration, 
if the Eulerian comoving energy density is positive \emph{anywhere}, 
then it must be negative \emph{somewhere else} --- so the WEC is certainly violated.

\subsection{Null energy condition}

Now consider the NEC: Take equation \eqref{E:bar-p2} and integrate over all space.
Note 
\begin{align}
\int_\R3  \nabla_a ( K n^a)\;\d x \d y \d z &= 
\int_\R3 \{\partial_t K +  \partial_i ( K v^i)\}\;\d x \d y \d z, 
\nonumber\\
&= \partial_t\int_\R3  K\;\d x \d y \d z  + \int_\R3   \partial_i ( K v^i)\;\d x \d y \d z, 
\nonumber\\
&= \partial_t\int_\R3  \partial_i v^i\;\d x \d y \d z  + \int_\R3   \partial_i ( K v^i)\;\d x \d y \d z, 
\nonumber\\
&= \partial_t \,0 + 0, 
\nonumber\\
&= 0.
\end{align}
Thence,
\begin{equation}
\int_\R3 \bar p (t,x,y,z) \;\d x \d y \d z = \int_\R3 \rho(t,x,y,z) \;\d x \d y \d z,
\end{equation}
and so
\begin{equation}
\int_\R3 \{\rho(t,x,y,z) +\bar p (t,x,y,z)\} \;\d x \d y \d z = 2\int_\R3 \rho(t,x,y,z) \;\d x \d y \d z \leq0.
\end{equation}
(We have already seen in the previous subsection that this last quantity is nonpositive.)

Accordingly, in this generic tractor/pressor/stressor beam configuration, if the quantity $(\rho+\bar p)$ is positive \emph{anywhere}, then it must be negative \emph{somewhere else} --- so the NEC is certainly violated. 

Now, given that the NEC is the weakest of all standard, classical, point-wise energy conditions, we have that all the other energy conditions will also be violated. This has to hold for all tractor/pressor/stressor configurations based on modifications of the generic Natário warp drive. Furthermore, this is completely in accord with what we saw happen for generic warp drive space-times~\cite{Santiago:2021}.

 \FloatBarrier
\addtocontents{toc}{\protect\vspace{-5pt}}
\section{Special Cases}\label{sec:special}

We now consider three special cases that link our tractor/pressor/stressor discussion back to various previous warp drive analyses~\cite{Alcubierre:1994,Natario:2001,Lobo:2004,Lobo:2017,Alcubierre:2017,Santiago:2021}. The connections between the envelope and profile functions of the generic Natário case described by equations~\eqref{E:beam1}--\eqref{E:beam3} and those appearing in these special cases is summarised in table~\ref{tab:specialcases}.
	
	\begin{table}[h!]
		\centering
		\begin{tabular}{| c | c | c | c |}
			\hline
			\textbf{Generic Natário} & \textbf{Modified Alcubierre} & \textbf{Zero Expansion} & \textbf{Zero Vorticity} \\ \hline
			\textbf{envelope} $k(t,z)$ & $0$ & $-\partial_z v$ & $\Phi$ \\ \hline 
			\textbf{envelope} $v(t,z)$ & $v$ & $v$ & $\partial_z \Phi$ \\ \hline 
			\textbf{profile} $h(u)$ & $0$ & $h$ & $2f'$ \\ \hline
			\textbf{profile} $f(u)$ & $f$ & $2(h+uh')$ & $f$ \\
			\hline 
		\end{tabular}
		\caption{A summary of the connection between the generic Natário metric, its envelope functions $k$ and $v$, and its profile functions $h$ and $f$ on the one hand, and the various functions appearing in the special cases considered section~\ref{sec:special}.}.
		\label{tab:specialcases}
	\end{table}

\subsection{Modified Alcubierre warp flow}     
 \enlargethispage{30pt}
For this particular special case we will assume the field to be oriented along a fixed direction, for convenience taken to be the $z$ direction. This corresponds to taking the flow field to be:
\begin{align}
v_x(t,x,y,z) &=  0,\\
v_y(t,x,y,z) &=   0,\\
v_z(t,x,y,z) &= v(t,z)\, f(x^2+y^2).
\end{align}
For this modified Alcubierre flow field the vorticity is
\begin{equation}
\vec \omega = \Big( \partial_y v_z, -\partial_x v_z, 0\Big) = \Big(y,-x,0\Big) \; 2 v(t,z) f'(x^2+y^2)
\end{equation}
and hence
\begin{equation}
\vec \omega\cdot \vec \omega = 4 \, v(t,z)^2 \, (x^2+y^2) \, [f'(x^2+y^2)]^2.
\end{equation}

Now, using the result that for the Alcubierre warp field $T_{zz} = 3\rho$, obtained in \cite{Alcubierre:2017}, a standard computation yields~\cite{Alcubierre:1994,Lobo:2017,Alcubierre:2017,Santiago:2021}:
\begin{equation}
\label{E:alcubierre-Tzz}
T_{zz}(t,x,y,z) = 3\, \rho(t,x,y,z) =  -{3\over32\pi}\{  (\partial_x v_z)^2 +  (\partial_y v_z)^2 \} 
= -{3\over8\pi} (\vec \omega\cdot \vec \omega) \leq 0.
\end{equation}
This is already enough to guarantee that both the weak energy condition (WEC) and null energy condition (NEC) are violated in this space-time~\cite{Alcubierre:2017,Santiago:2021}. 
Calculating the net force we obtain for a narrow beam: 
\begin{equation}
\label{E:alcubierre-F}
F(t,z) =   -{3\over32\pi} \int_\R2 [ (\partial_x v_z)^2 + (\partial_y v_z)^2 ] \; \d x \d y< 0.
\end{equation}
But, given our factorization assumption, the stress reduces to 
\begin{equation}
T_{zz}(t,x,y,z) =  -{3\over32\pi} v(t,z)^2 \{  (\partial_x f)^2 +  (\partial_y f)^2 \}  < 0.
\end{equation}
Under this assumption the force factorizes to
\begin{equation}
F(t,z) =   -{3\over32\pi} v(t,z)^2 \int_\R2 [ (\partial_x f)^2 + (\partial_y f)^2 ]\; \d x \d y< 0.
\end{equation}
That is, using $u=x^2+y^2$,
\begin{equation}
F(t,z) =   -{3\over8}\; v(t,z)^2 \int_0^\infty  u [f'(u)]^2\; \d u < 0.
\end{equation}
This is always a tractor beam. 
The $x$-$y$ integral is just some positive dimensionless number characterizing the shape of the beam.
(Recall that our convention was to always put the target to the right of the generator. If we flip target and generator, so that the target is now on the left and the beam impinges on the target from the right, then there is a sign flip for the force $F(t,z)$, and with $F(t,z) >0$ the target is still attracted to the generator.)

If we instead assume a wide beam, one can immediately deduce that in this case equation~\eqref{E:generic-wide} will always reduce to zero, as either $k$ or $h$ is zero.

\subsection{Zero-expansion beam}  
\enlargethispage{40pt}

Now consider a zero-expansion  flow field subject to $\partial_i v^i=0$. Starting with the generic flow field appropriate to an axisymmetric beam, we have:
\begin{align}
	v_x(t,x,y,z) &=  k(t,z)\, x \,  h(x^2+y^2)\\
	v_y(t,x,y,z) &=   k(t,z)\, y \,  h(x^2+y^2)\\
	v_z(t,x,y,z) &= v(t,z)\, f(x^2+y^2).
\end{align}
Then, in order to ensure zero expansion, we must enforce
\begin{equation}
\partial_i v^i = \partial_z v(t,z)\; f(x^2+y^2) + 2 k(t,z)\; \{h(x^2+y^2) +  (x^2+y^2) h'(x^2+y^2) \} = 0.
\end{equation}
Separating variables, one finds
\begin{equation}
{ \partial_z v(t,z) \over k(t,z)} = - { 2[h(x^2+y^2) + (x^2+y^2) h'(x^2+y^2)]\over f(x^2+y^2)} ={C},
\end{equation}
for some separation constant $C$. 

Then, without loss of generality, we can enforce:
\begin{equation}
k(t,z) = - \partial_z v(t,z), \qquad f(x^2+y^2) = 2[h(x^2+y^2) + (x^2+y^2) h'(x^2+y^2)].
\end{equation}

Therefore, the zero-expansion flow field can be rewritten in terms of only two free functions $v(t,z)$ and $h(x^2+y^2)$:
\begin{align}
	\label{E:zero-vx}
		v_x(t,x,y,z) &=  -x \, \partial_z v(t,z) \,h(x^2+y^2),\\
	\label{E:zero-vy}
		v_y(t,x,y,z) &=  -y \,  \partial_z v(t,z) \,h(x^2+y^2),\\
	\label{E:zero-vz}
		v_z(t,x,y,z) &= \hphantom{-}2 v(t,z)\; \{ h(x^2+y^2) + (x^2+y^2) h'(x^2+y^2)  \} 
\end{align}
This flow field automatically satisfies axial symmetry, a beam-like profile,  and zero expansion.
So this is indeed suitable for describing a zero expansion ``beam''. The vorticity for this beam is easily evaluated as
\begin{equation}
	\vec \omega = \Big(y,-x,0\Big) 
	\left\{ \partial_z^2 v(t,z) h(u) + 4v(t,z)[u^2 h'(u) ]'/u\right\}.
\end{equation}
Thence,
\begin{equation}
	\vec \omega \cdot \vec \omega =  u\,
	\Big\{ \partial_z^2 v(t,z) h(u) + 4v(t,z)[ u^2 h'(u)]'/u\Big\}^2.
\end{equation}

\subsubsection{Force}  
Let us now calculate $T_{zz}(t,x,y,z)$ for this flow field,  and then integrate over the $x$-$y$ plane in order to obtain the net force.
For $T_{zz}(t,x,y,z)$ we find:
\begin{align}
	T_{zz}(t,x,y,z) =& {1\over 8\pi} \left\{ Z_1(x,y) \;v(t,z) \;\partial_z^2 v(t,z) + Z_2(x,y)\,[\partial_z^2 v(t,z) ]^2  
	+ Z_3(x,y) [\partial_z v(t,z)]^2 
	\right.\nonumber\\
	& \qquad \left.
	 + Z_4(x,y) \;\partial_t \partial_z v(t,z) + Z_5(x,y) \; v(t,z)^2\right\}.
\end{align}
Again using the shorthand $u=x^2+y^2$, we can explicitly calculate:
\begin{align}
	\label{E:Z1}
	Z_1(x,y) &=\hphantom{-}[u^2 h(u)^2]'' + 2 ([u h(u)]')^2 ,\\
	\label{E:Z2}
	Z_2(x,y) &=\hphantom{-} {1\over4} u h(u)^2 ,\\
	\label{E:Z3}
	Z_3(x,y) &= - 2[u^2 h(u)^2]'' + [uh(u)^2]',\\
	\label{E:Z4}
	Z_4(x,y) &= \hphantom{-}2 [u h(u)]',\\
	\label{E:Z5}
	Z_5(x,y) &= -\frac{12}{u}\left([u^2h'(u)]'\right)^2 .
\end{align}
Now consider the integrals over the $x$-$y$ plane. But first note that
\begin{equation}
	\int_\R2 Z_i(x,y) \, \d x \d y= \pi \int_0^\infty Z_i(u)\d u.
\end{equation}
Because you want the beam to die off far away from the beam axis, you want $h(x^2+y^2)= h(u) \to 0$ as $x^2+y^2= u\to\infty$. 
So we can already extract some limited information regarding the integrals:
\begin{gather}
	\int_\R2 Z_1(x,y) \; \d x \d y \; >0 , \quad \int_\R2 Z_2(x,y) \; \d x \d y>0,\\
	\int_\R2 Z_3(x,y) \; \d x \d y=0, \\
	\int_\R2 Z_4(x,y) \; \d x \d y=0,\\
	\int_\R2 Z_5(x,y) \; \d x \d y<0.
\end{gather}

Overall, for the zero-expansion narrow beam we now have
\begin{align}
	F(t,z) =& {1\over 8} \left\{ 2 v(t,z) \;\partial_z^2 v(t,z)  \int_0^\infty ([u h(u)]')^2 du 
	+ {1\over 4}\,[\partial_z^2 v(t,z) ]^2   \int_0^\infty u h(u)^2 du 
	\right.\nonumber\\
	& \qquad \left.
	- 12 \; v(t,z)^2 \int_0^\infty \left([u^2h'(u)]'\right)^2 \d u\right\}.
\end{align}
The first term is indefinite (even though the coefficient is positive), the second term is positive semi-definite, and the third term is negative semi-definite. So the zero-expansion narrow beam can be tuned to be either a tractor or a pressor, (or even a stressor). One cannot say more about the force $F(t,z)$ without making a specific choice for the profile $h(u)$, and the envelope function $v(t,z)$.

\enlargethispage{10pt}
If we now consider a wide beam, then  we should look on axis and evaluate $T_{zz}(t,0,0,z)$. Specifically, we see:
\begin{gather}
	Z_1(0,0) = 4 h(0)^2, \quad Z_2(0,0) = 0, \quad Z_3(0,0) = -3h(0)^2, \nonumber\\
	Z_4(0,0) = 2h(0) \quad
	Z_5(0,0) = 0. \qquad \qquad\quad
\end{gather}
whence,
\begin{equation}
	\label{eq: zero exp axis}
	T_{zz}(t,0,0,z) = \frac{1}{8\pi}\left\{
	4h(0)^2 [v(t,z)\partial_z^2 v(t,z)] -3h(0)^2[\partial_z v(t,z)]^2 + 2h(0)[\partial_t\partial_z v(t,z)]
	\right\}.
\end{equation}
So in the wide beam limit of a zero-expansion beam we have
\begin{equation}
	F(t,z)= \frac{h(0)}{8\pi}\left\{
	4h(0) [v(t,z)\partial_z^2 v(t,z)] -3h(0)[\partial_z v(t,z)]^2 + 2[\partial_t\partial_z v(t,z)]
	\right\} A.
\end{equation}	

\subsubsection{Energy conditions}  

For this zero expansion space-time we have $K = \tr(K_{ij})=0$, and so from equations \eqref{E:rho} and \eqref{E:bar-p}  it is immediate that
\begin{equation}
	\rho = \bar p = - {1\over16\pi} \tr(K^2)  <0.
\end{equation}
This is enough to guarantee that both the WEC and NEC are violated~\cite{Santiago:2021},
but for the sake of completeness we perform an explicit calculation.

\paragraph{WEC:} For the Eulerian energy density we note: 
\begin{align}
	\rho(t,x,y,z) =& {1\over 8\pi} \left\{ 
	W_1(x,y) \;v(t,z) \;\partial_z^2 v(t,z) 
	+ W_2(x,y)\,[\partial_z^2 v(t,z) ]^2  
	\right.\nonumber\\
	& \qquad \left.
	+ W_3(x,y) [\partial_z v(t,z)]^2 
	 + W_4(x,y) \; v(t,z)^2\right\}.
\end{align}
Using the shorthand $u=x^2+y^2$, we can explicitly calculate:
\begin{align}
	W_1(x,y) &= \hphantom{-}2h(u)[u^2h'(u)]',\\
	W_2(x,y) &= -{1\over4} u h(u)^2,\\
	W_3(x,y) &= -4 ( [u h(u)]')^2 + [u h(u)^2]',\\
	W_4(x,y) &= -\frac{4}{u}\{[u^2h'(u)]'\}^2.
\end{align} 

This is almost a (negative) perfect square:
\begin{eqnarray}
\rho(t,x,y,z)  &=& - {1\over8\pi} u \left[  {h(u)\over2} \partial_z^2  v(t,z)- {2\over u}   v(t,z)  {[u^2h'(u)]'} \right]^2 
\nonumber\\
&&+{1\over8\pi} \{-4([uh(u)]')^2 +[uh(u)^2]'\} [\partial_z v(t,z)]^2.
\end{eqnarray}
By performing an integration over the $x$-$y$ plane, this can then be fully written as the sum of negative perfect squares:
\begin{align}
	\int_\R2\rho(t,x,y,z)\d x \d y  = -& {1\over8} \int_{0}^{\infty}u \left[  {h(u)\over2} \partial_z^2  v(t,z)- {2\over u}   v(t,z)  {[u^2h'(u)]'} \right]^2 \d u \nonumber \\
	-&{1\over2}\int_{0}^{\infty}([uh(u)]')^2 \d u  \; [\partial_z v(t,z)]^2\leq 0.
\end{align}
This is more than sufficient to guarantee WEC violation somewhere on each $x$-$y$ plane. Now let us also integrate over $\d z$. Using $ \int  [\partial_z v(t,z)]^2 \d z = - \int  v(t,z) \partial_z^2 v(t,z) \d z$, we obtain:
\begin{align}
	\int_\R3\rho(t,x,y,z)\d x \d y  \d z = -& {1\over8} \int_{0}^{\infty}u \left[  {h(u)\over2} \partial_z^2  v(t,z)- {2\over u}   v(t,z)  {[u^2h'(u)]'} \right]^2 \d u \d z \nonumber \\
	+&{1\over2}\int_{0}^{\infty}([uh(u)]')^2 \d u  \; \int v(t,z) [\partial_z^2 v(t,z)] \d z .
\end{align}
Now, note that
\begin{align}
([uh(u)]')^2 &= (h(u) + u h'(u))^2 = h(u)^2 + 2 u h(u)h'(u) + u^2 h'(u)^2 
\\
&= [u h(u)^2]' +u^2 h'(u)^2
= [u h(u)^2]' +  [u^2 h(u) h'(u) ]' - h(u) [u^2 h'(u)]'.\qquad \nonumber
\end{align}
So, after an integration by parts,
\begin{equation}
\int_{0}^{\infty}([uh(u)]')^2 \d u  = - \int_{0}^{\infty}h(u) [u^2 h'(u)]' \d u.
\end{equation}
But this now implies that we have a (negative) perfect square:
\begin{equation}
	\int_\R3\rho(t,x,y,z)\d x \d y  \d z = - {1\over8} \int_{0}^{\infty}u \left[  {h(u)\over2} \partial_z^2  v(t,z)+ {2\over u}   v(t,z)  {[u^2h'(u)]'} \right]^2 \d u \d z \leq 0.
\end{equation}
Even more, we can recognize this as
\begin{equation}
	\int_\R3\rho(t,x,y,z)\d x \d y  \d z = - {1\over32} \int_{0}^{\infty}
	(\vec\omega\cdot\vec\omega)  \;\d u \d z \leq 0,
\end{equation}
as expected. Again, this is more than sufficient to guarantee WEC violation somewhere on each spatial slice, apart from also verifying internal consistency of the formalism.

\paragraph{NEC:} To prove the violation of the NEC, we must now look at the quantity $\left[\rho + T_{zz}\right]$:
\begin{align}
	\left[\rho + T_{zz}\right](t,x,y,z) =& {1\over8\pi} 
	\Big\{ X_1(x,y) [v(t,z)\partial_z^2v(t,z)]
	+
	X_2(x,y)[\partial_z v(t,z)]^2
	\nonumber\\
	& \qquad
	+
	X_3(x,y) [v(t,z)]^2\Big\},
\end{align}
with \enlargethispage{30pt}
\begin{align}
	X_1(x,y) &= 2 [u^2 h(u)^2 ]'',\\
	X_2(x,y) &= -4 [u h'(u)]^2 - 2 [ u^2 h(u)^2]'' - 2 [u h(u)^2]',\\
	X_3(x,y) &= -16 u [u h''(u)+ 2 h'(u)]^2.
\end{align}

Consequently, integrating over the $x$-$y$ plane we have:
\begin{align}
	\int_\R2  \left(\rho + T_{zz}\right) \d x \d y =& \hphantom{-}\pi \int_0^\infty  \left(\rho + T_{zz}\right)  du 
	\nonumber\\
	=& 
	 -{1\over2}  [\partial_z v(t,z)]^2 \int_0^\infty [u h'(u)]^2 \d u\; 
	 \nonumber\\
	&-2 [v(t,z)]^2 \int_0^\infty u [u h''(u)+ 2 h'(u)]^2\d u\;  .
\end{align}
This is now a (negative) sum of squares, thereby guaranteeing violation of the NEC. 
This is again a useful consistency check on the formalism.

\subsection{Zero-vorticity beam}  
 Let us now consider a zero-vorticity beam described by the flow field:
\begin{equation}
	v_i(t,x,y,z) =\partial_i \tilde{\Phi}, \qquad \tilde{\Phi}(t,x,y,z) \to \Phi(t,z) f(x,y).
\end{equation}
The stress component $T_{zz}(t,x,y,z)$ will again be somewhat complicated. 
However once one integrates over the $x$-$y$ plane we shall soon see that 
\begin{equation}
	F(t,z) =  \int_\R2  T_{zz} \;  \d x \d y = 0.
\end{equation}
That is, there is no \emph{net} force once you integrate over the entire 2-plane. We shall soon see, however, that there are regions of both repulsion/attraction at various points on the 2-plane. This is best interpreted as  a stressor beam.

\subsubsection{Force}

Explicitly calculating  the stress component $T_{zz}(t,x,y,z)$ we find:
\begin{equation}
	T_{zz}(t,x,y,z) = {1\over8\pi} \{ P_1(x,y) \; \partial_t \Phi(t,z) 
	+ P_2(x,y) \; (\partial_z \Phi(t,z))^2 + P_3(x,y) \; \Phi(t,z)^2\}.
\end{equation}
Now, using again $u=x^2+y^2$ for compactness, explicit computation yields
\begin{align}
	P_1(x,y) &= - 4 [u f'(u)]',\\
	P_2(x,y) &= - 4 [u f(u) f'(u)]',\\
	P_3(x,y) &= - [\{16 u^2 f'(u)f''(u)\} +12 u [f '(u)]^2     ]'.
\end{align}
Noting again that
\begin{equation}
	\int_\R2 P_i(x,y)  \d x \d y= \pi \int_0^\infty P_i(u)\d u,
\end{equation}
and observing that each of the $P_i$ is a pure derivative, one has
\begin{equation}
	\int_\R2 P_i(x,y)  \d x \d y= \pi \int_0^\infty P_i(u)\d u = 0.
\end{equation}
So there is no \emph{net} force.

Note, however, that:
\begin{equation}
	T_{zz}(t,0,0,z) = -{f'(0)\over2\pi} \{  \partial_t \Phi(t,z) 
	+ f(0) \; (\partial_z \Phi(t,z))^2  + 3 f '(0) \; \Phi(t,z)^2\},
\end{equation}
which will in general not equal zero. In this way we see that, while the force integrated all over the $x-y$ plane sums up to zero, this does not imply an identically zero force. On the contrary, different parts of the target will be pulled, while others pushed, creating a perfect example of a stressor beam. Furthermore, this means that---ignoring matters of material properties---a wide zero-vorticity beam could potentially still act as a tractor or pressor beam, whereas a narrow zero-vorticity beam would not.

\subsubsection{Energy density and null energy condition}\label{SS:ec+null}

Calculating the Eulerian (comoving) energy density we find:
\begin{equation}
	\rho(t,x,y,z) = {1\over 8\pi} \left\{ R_{zz}\; \Phi(t,z)[\partial_z^2 \Phi(t,z) ]  + R_{z} \;[\partial_z \Phi(t,z)]^2  + R_{0} \; \Phi(t,z)^2\right\}.
\end{equation}
with 
\begin{align}
	R_{zz} &= \hphantom{-}4 f(u) [u f''(u) + f'(u)] = 4 f(u)[u f'(u)]',\\[2pt]
	R_0 &= \hphantom{-}8u f''(u) f'(u) + 4 (f'(u))^2 = 4[u (f'(u))^2]',\\[2pt]
	R_z &= -4u (f'(u))^2.
\end{align}

Note that $\int_\R2 R_0 \; \d x \d y=0$, whereas after an integration by parts:
\begin{equation}
	\int_\R2 R_z \; \d x \d y= \int_\R2 R_{zz} \; \d x \d y= - 4\pi  \int_0^\infty u [f'(u)]^2\d u.
\end{equation}
Hence,
\begin{equation}
	\int_\R2 \rho \; \d x \d y = -  {1\over2} 
	\{  \Phi(t,z)[\partial_z^2 \Phi(t,z) ] + [\partial_z \Phi(t,z)]^2 
	  \} \int_0^\infty u [f'(u)]^2\d u.
\end{equation}
That is,
\begin{equation}
	\int_\R2 \rho \; \d x \d y = - {1\over2}   \; \partial_z \{ \Phi(t,z)\partial_z \Phi(t,z) \} \int_0^\infty u [f'(u)]^2\d u .
\end{equation}
Now, given the fall-off conditions on $\Phi(t,z)$, namely that $v(t,z) =\partial_z\Phi(t,z) \stackrel{z\to \pm \infty}{\longrightarrow} 0$, we have that:
\begin{equation}
	\label{E:zero vort rho}
	\int_\R3 \rho \; \d x \d y \d z =  0.
\end{equation}
Therefore, if the zero vorticity stressor beam has positive density anywhere, then it must have negative energy density somewhere else.  Thence, this zero-vorticity configuration violates the WEC. This is fully in agreement with the general warp-drive analysis presented in~\cite{Santiago:2021}.
\enlargethispage{20pt}

Furthermore, since we have already seen $\int_\R2 T_{zz} \, \d x\, \d y=0$, it automatically follows that  
$\int_\R3 T_{zz} \; \d x \d y \d z=0$, 
and thence we have $\int_\R3 (\rho+T_{zz}) \; \d x\, \d y\, \d z=0$. 
So, just like before, if the zero vorticity stressor  beam has $(\rho+T_{zz})$ positive anywhere, then this quantity must be negative somewhere else.  Therefore this zero-vorticity configuration also violates the NEC. Again, this is fully in agreement with the general warp-drive analysis presented in~\cite{Santiago:2021}, and is a useful consistency check on the fact that zero-vorticity flow fields do indeed violate the NEC.

\FloatBarrier
\addtocontents{toc}{\protect\vspace{-5pt}}
\section{Specific Examples}

In lieu of direct knowledge how one would actually build a tractor beam, one is left with two extremes: General considerations or modelling of specific possibilities. Our discussion in sections \ref{sec: Natario} and \ref{sec:Beam profile} are based on the generic Natário warp drive, in a sense a compromise between the two. While fixing, for example, a certain $(3+1)$ split and flat spatial slices in this split, it still retains a large amount of freedom. Section~\ref{sec:special} then considered more constrained choices found in the literature, while still retaining some freedom to choose certain functions appearing therein. In this section, we shall illustrate the results of the preceding sections for specific profile functions $f$ and $h$ and specific envelope functions $k, v$. 
As a first step, we shall start by imposing a Gaussian profile by fixing the functions $f(x^2 + y^2)$ and $h(x^2 + y^2)$ to be Gaussian functions. In a second step, we will then employ envelope functions that contribute to the stress-energy tensor between the positions of the generator (at $z_{\text{generator}}\ll 0$) and the target (at $z_{\text{target}}>z_{\text{generator}}$), while vanishing exactly outside of some certain region $(-b,b)$ on the $z$-axis. More specifically, we will adopt a kind of smooth \enquote{bump function}. Naturally, these are by far not the only choices, and they contain a certain amount of arbitrariness. Nevertheless, this should give a good idea of what can be done if an arbitrarily advanced civilization could impose stress-energy sources in such a targeted way.

\subsection{Gaussian beam profiles}
\enlargethispage{20pt}
As Gaussian beam profiles are very popular toy models in optics and acoustics, they are an obvious starting point for investigating our tractor beams. Let us then provide a few specific examples based on Gaussian beam profiles in the following discussion.

\subsubsection{Generic Gaussian beam}\label{sec:Gauss}

Let us consider a generic Gaussian beam, where we set the two profile functions  to be identical Gaussians with width parameters $a$:
\begin{align}
	v_x(t,x,y,z) &=  x \, k(t,z)\, \exp(-[x^2+y^2]/a^2),\\
	v_y(t,x,y,z) &=  y \,  k(t,z)\, \exp(-[x^2+y^2]/a^2),\\
	v_z(t,x,y,z) &= v(t,z)\, \exp(-[x^2+y^2]/a^2).
\end{align}
Then, based on equations \eqref{E:generic-Tzz} and \eqref{E:generic-N1}--\eqref{E:generic-N5}, we see:
\begin{align}
	N_1(x,y) &= - {3 u\over a^{4}}  \exp(-2u/a^2 ) ,\\
	N_2(x,y) &= \hphantom{-}{(3u-2a^2)\over a^2}  \exp(-2u/a^2), \\
	N_3(x,y) &= \hphantom{-}{u\over4}   \exp(-2u/a^2 ) ,\\
	N_4(x,y) &=  \hphantom{-}  {2(u-a^2)\over a^2}  \exp(-u/a^2 ) ,\\
	N_5(x,y) &= - {(4u-a^2)(2u-3a^2)\over a^4}  \exp(-2u/a^2 ).
\end{align}
Now use
\begin{equation}
	\int_\R2 N_i(x,y) \, \d x \d y= \pi \int_0^\infty N_i(u)\d u,
\end{equation}
to get
\begin{gather}
	\int_\R2 N_1(x,y) \, \d x \d y=  - {3\pi\over4},\qquad
	\int_\R2 N_2(x,y) \, \d x \d y= - {1\over 4} \pi a^2, \\
	\int_\R2 N_3(x,y) \, \d x \d y=  \pi{a^4\over16},\qquad 
	\int_\R2 N_4(x,y)\,  \d x \d y=  0, \qquad
	\int_\R2 N_5(x,y) \, \d x \d y=  0.
\end{gather}
Then, for the net force exerted by this Gaussian beam (in the narrow field limit), we find a particularly  simple factorized form:
\begin{align}
	F(t,z) &=  -\frac{F_*}{32\; c^2} \left\{   {3} {v(t,z)^2} 
	+{a^2} {v(t,z) \partial_z k(t,z) } 
	- {a^4 [\partial_z k(t,z)]^2\over 4 }\right\} \nonumber\\
	 &=  -\frac{3}{32} 
	\left\{ v(t,z) -{1\over6} a^2 \, \partial_z k(t,z)  \right\} 
	\left\{ v(t,z) +{1\over2} a^2 \, \partial_z k(t,z) \right\} .
\end{align}
Note that the behaviour switches from pressor to tractor when the beam satisfies  the two critical conditions:
\begin{equation}
 v(t,z) ={1\over6} a^2 \, \partial_z k(t,z), \qquad 
 v(t,z) =-{1\over2} a^2 \, \partial_z k(t,z). 
\end{equation}
So, adjusting the two envelope functions is the determining factor in choosing tractor/\-pressor/\-stressor behaviour.

In contrast, on axis we see that
\begin{equation}
N_1(0,0) = 0, \quad N_2(0,0) = -2, \quad  N_3(0,0) = 0, \quad  N_4(0,0) = -2, \quad 
N_5(0,0) = -3.\qquad 
\end{equation}

So that 
\begin{equation}
\label{E:gaussian-axis--Tzz}
	T_{zz}(t,0,0,z) = -{1\over8\pi} \left\{2 \;v(t,z) \partial_z k(t,z) 
	+2 \partial_t k(t,z)+ 3 \; k(t,z)^2 \right\}.
\end{equation}
So, in the wide-beam limit
\begin{equation}
\label{E:f-gaussian-wide-beam}
	F(t,z) = -{1\over8\pi} \left\{2 \;v(t,z) \partial_z k(t,z) 
	+2 \partial_t k(t,z)+ 3 \; k(t,z)^2 \right\} A.
\end{equation}
This is of indefinite sign, depending delicately on the envelope functions, potentially allowing the wide-beam to be fine-tuned as either a tractor or a pressor. In SI units
\begin{equation}
\label{E:f-gaussian-wide-beam2}
	F(t,z) = -{F_*\over8\pi c^2} \left\{2 \;v(t,z) \partial_z k(t,z) 
	+2 \partial_t k(t,z)+ 3 \; k(t,z)^2 \right\} A.
\end{equation}
We remind the reader that the Stoney force is (by human standards) truly enormous,  $F_* = c^4/G_\mathrm{Newton} \approx 1.2 \times 10^{44}$ Newtons. 
Even relatively small envelope functions $v(t,z)$ and $k(t,z)$ can lead to significant tractor/pressor effects.

\enlargethispage{40pt}
\subsubsection{Alcubierre-based  Gaussian beam}
Let us now consider a Gaussian beam based on the modified Alcubierre flow field. Take $f(x,y) = \exp(-[x^2+y^2]/a^2)$, then from \eqref{E:alcubierre-Tzz} and \eqref{E:alcubierre-F} we ultimately see
\begin{equation}
	\int_\R2 [ (\partial_x f)^2 + (\partial_y f)^2 ] \; \d x \d y\to \int_\R2 {4(x^2+y^2)\over a^4} \exp(-[x^2+y^2]/a^2) \; \d x \d y=   \pi.
\end{equation}
Note this Gaussian profile implies that $T_{zz}(t,x,y,z)$ is zero on the $z$ axis, rises to a maximum for $(x^2+y^2)\sim a^2$, and then very rapidly decays as you move further off axis. For the total net force on the $x$-$y$ plane this Gaussian beam gives:
\begin{equation}
	F(t,z) \to   -{ 3\over32} \; v_0(t,z)^2. 
\end{equation}

Putting back all the appropriate dimensions, we obtain in SI units
\begin{equation}
	F(t,z) \to   - {3 F_*\over32} \;  {v_0(t,z)^2 \over c^2}.
\end{equation}
Here $F_*$ is again the Stoney force. Note that, as expected, this is always a tractor beam.

\subsubsection{Zero-expansion Gaussian beam}
Looking now at a zero-expansion Gaussian beam, we set $h(x^2+y^2) = \exp(-(x^2+y^2)/a^2)$. Then, using \eqref{E:zero-vx}--\eqref{E:zero-vy}--\eqref{E:zero-vz} and \eqref{E:Z1}--\eqref{E:Z5}, we have
\begin{align}
	Z_1(x,y)&= 4 \{ 1 -3 u a^{-2} + 3 u^2 a^{-4} /2\}  \exp(-2u/a^2),\\
	Z_2(x,y)&= {1\over4} u \exp(-2u/a^2),\\
	Z_3(x,y)&= \{-3 +14u/a^2-8u^2/a^4\}\exp(-2u/a^2),\\
	Z_4(x,y)&= 2 \{1-u/a^2 \}\exp(-u/a^2),\\
	Z_5(x,y)&= -12 u (2-u/a^2)^{2} a^{-4} \exp(-2u/a^2).
\end{align}
Thence, for the relevant integrals
\begin{gather}
	\int_\R2 Z_1(x,y)\;  \d x \d y= \pi \;{a^2\over2}, \qquad
	\int_\R2 Z_2(x,y) \; \d x \d y= \pi \;{a^4\over16}, \qquad
	\int_\R2 Z_3(x,y) \; \d x \d y= 0, \nonumber\\
	\int_\R2 Z_4(x,y) \; \d x \d y= 0, \qquad
	\int_\R2 Z_5(x,y) \; \d x \d y= -\pi \;{9\over2}. \qquad \;
\end{gather}
So, for the Gaussian zero expansion beam, we see that
\begin{equation}
	F(t,z) = {1\over 8} \left\{ {a^2\over2} v(t,z) \partial_z^2 v(t,z) + {a^4\over 16} [\partial_z^2 v(t,z) ]^2  
	-{9\over2}  v(t,z)^2\right\}.
\end{equation}
This can be either be a pressor or a tractor beam, depending on the choice of the envelope function.

Now consider the wide beam limit. For a Gaussian zero-expansion beam equation \eqref{eq: zero exp axis} for $T_{zz}$ reads:
\begin{equation}
T_{zz}(t,0,0,z) = \frac{1}{8\pi}\left\{
4 [v(t,z)\partial_z^2 v(t,z)] -3[\partial_z v(t,z)]^2 + 2[\partial_t\partial_z v(t,z)] 
\right\}.
\end{equation}
In SI units,
\begin{equation}
T_{zz}(t,0,0,z) = \frac{F_*}{8\pi c^2}\left\{
4 [v(t,z)\partial_z^2 v(t,z)] -3[\partial_z v(t,z)]^2 + 2[\partial_t\partial_z v(t,z)] 
\right\} A.
\end{equation}
Again, this can be either be a pressor or a tractor beam, depending on the choice of the envelope function.

\subsubsection{Zero vorticity Gaussian beam}
If we now take a specific Gaussian profile $f(x^2 + y^2) = \exp(-[x^2+y^2]/a^2)$ then, for a zero vorticity beam we find:
\begin{align}
	P_1(x,y) &= -4 (x^2+y^2-a^2) a^{-4} \exp(-[x^2+y^2]/a^2),\\
	P_2(x,y) &= -4 (2[x^2+y^2]-a^2) a^{-4} \exp(-2[x^2+y^2]/a^2),\\
	P_3(x,y) &= -4 (2[x^2+y^2]-3a^2)  (4[x^2+y^2]-a^2) a^{-8} \exp(-2[x^2+y^2]/a^2).
\end{align}
We can explicitly check that
\begin{equation}
\int_\R2 P_i (x,y) \d x \d y = 0.
\end{equation}
The sign of the $P_i(x,y)$ and consequently the sign of $T_{zz}(t,x,y,z)$ can and will change near $x^2+y^2 \sim a^2$, so the spatially target will be alternately pushed and pulled --- which is why we classify this case as a stressor beam. 
The calculation on-axis ($x=y=0$) gives us:
\begin{align}
	P_1 \to& \hphantom{-}4 a^{-2}, \nonumber\\
	P_2 \to&  \hphantom{-} 4 a^{-2} ,\\
	P_3 \to&  -12 a^{-4} , \nonumber
\end{align}
which results in:
\begin{equation}
	T_{zz}(t,0,0,z) =  {1\over8\pi} \{4 a^{-2} \; \partial_t \Phi(t,z) + 4 a^{-2} \; (\partial_z \Phi(t,z))^2 -12 a^{-4} \; \Phi(t,z)^2\} .
\end{equation}
So, in the wide-beam limit,
\begin{equation}
F(t,z) = {1\over8\pi} \{4 a^{-2} \; \partial_t \Phi(t,z) + 4 a^{-2} \; (\partial_z \Phi(t,z))^2 -12 a^{-4} \; \Phi(t,z)^2\} A.
\end{equation}
As we can see, this is another \enquote{tunable} case, which can behave either as a pressor or a tractor beam, depending on the choice of the envelope function.

\FloatBarrier
\enlargethispage{30pt}
\subsection{Envelope functions}
In order to be able to visualize some of the properties of tractor/pressor/stressor beams we shall now impose two different possibilities for the envelope functions $v(t,z)$ and $k(t,z)$. This will allow us to plot the force field generated by these functions and the energy density distribution necessary to create them. All of the calculations done in the previous sections are still completely valid here.

\subsubsection{Illustrating Gaussian beams}\label{sec:GaussEnv}

In figure~\ref{fig:Gauss}, used in the Introduction to describe where target and generator are with respect to the tractor field, we also plotted the energy densities and forces of the non-trivial beam configurations described above. To produce those plots, we imposed a Gaussian envelope together with a Gaussian profile for the defining functions:
\begin{align}
	f_{\text{Gauss, plot}} = h_{\text{Gauss, plot}}&= e^{-u/A^2},\nonumber\\
	v_{\text{Gauss, plot}} = k_{\text{Gauss, plot}} = \Phi_{\text{Gauss, plot}}&= e^{-z^2/B^2}e^{-t^2/C^2},\label{E:GaussPlot}
\end{align}
where, for the plotting, we used $A=0.5$, $B=C=1.0$, and we evaluated the energy density and forces at $t=1$. Note that $F(t,z)$ for both the narrow Alcubierre and the wide zero-vorticity beams are always negative for this specific setup, implying a tractor beam behaviour, while the other beam configurations allow for a tractor/pressor behaviour depending on the positioning of the target. It is also nice to notice how non-trivial is the cancelation of the energy density along the spatial 3-slices for the zero vorticity case, given by equation \eqref{E:zero vort rho} and represented in figure~\ref{fig:Gauss}-(b).

\subsubsection{Bump functions}

\begin{figure}
	\centering
	\includegraphics[scale=0.65]{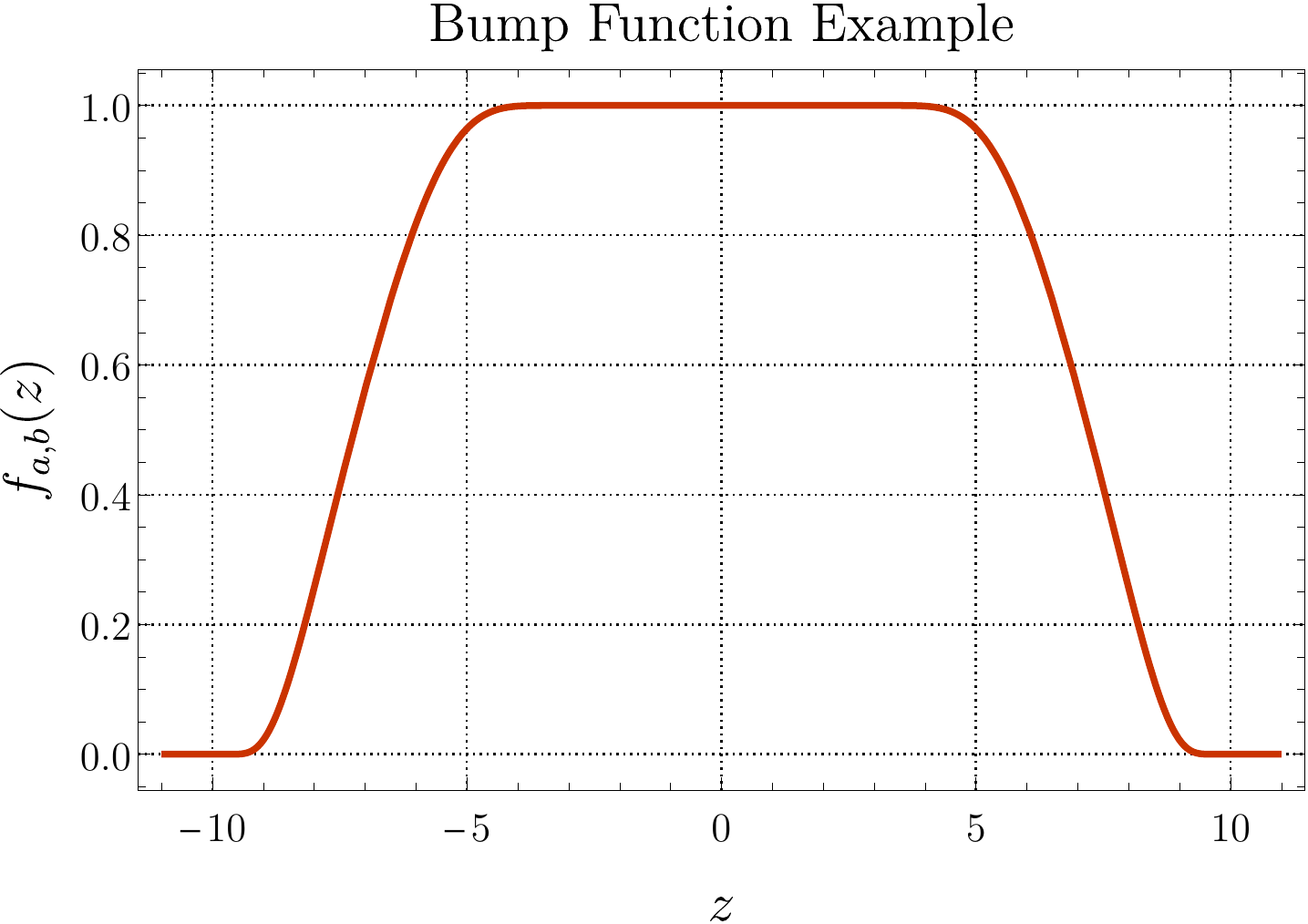}
	\caption{An illustration of the bump function given by equation \eqref{E:Bump2}, with $a=2$ and $b=10$.}
\label{F:bump}
\end{figure}

Another, much more brutal way of enforcing that fall-off conditions be fulfilled is by using smooth bump functions \cite[\S 13.1]{Tu}. Concretely, we will employ the following examples of smooth functions of compact support: First, define
\begin{equation}
	f_1(z) = \begin{cases}
		e^{-1/z} & z > 0\\
		0 & \text{else}.
	\end{cases}
\end{equation}
Use this to then define
\begin{equation}
	f_2(z) = \frac{f_1(z)}{f_1(z)+f_1(1-z)}.
\end{equation}
In a last step, define for real numbers $a$ and $b$
\begin{equation}
	\label{E:Bump2}
	f_{a,b}(z) = 1 - f_2\left( \frac{z^2-a^2}{b^2-a^2} \right).
\end{equation}
This function is $0$ for $z \in (-\infty,-b) \cup (b,\infty)$, is $1$ in the interval $(-a,a)$, smoothly grows from $0$ to $1$ on $[-b,-a]$ and decays smoothly from $1$ to $0$ on $[a,b]$, as can be seen in figure~\ref{F:bump}.

\bigskip
As we are interested in functions satisfying appropriate fall-off conditions at infinity, this example fulfils this by construction in the most trivial way possible: It vanishes for sufficiently large positive or negative values of $x$. Furthermore, as we are specifying the metric by hand, the Einstein equation will tell us the required sources; just as in all the calculations of this paper. Neither the Gaussian beams nor beams based on such smooth bump functions differ in this regard from each other, and the general analysis of the previous sections will still hold. Nevertheless, using such smooth bump functions for the envelope functions $v$ or $k$ is an intriguing way to model a tractor beam that only contributes to the stress energy on the $z$-axis between \enquote{generating device} and \enquote{target}.

\begin{figure}
 	\centering
 	\begin{subfigure}[t]{\textwidth}
 		\includegraphics[width=\textwidth]{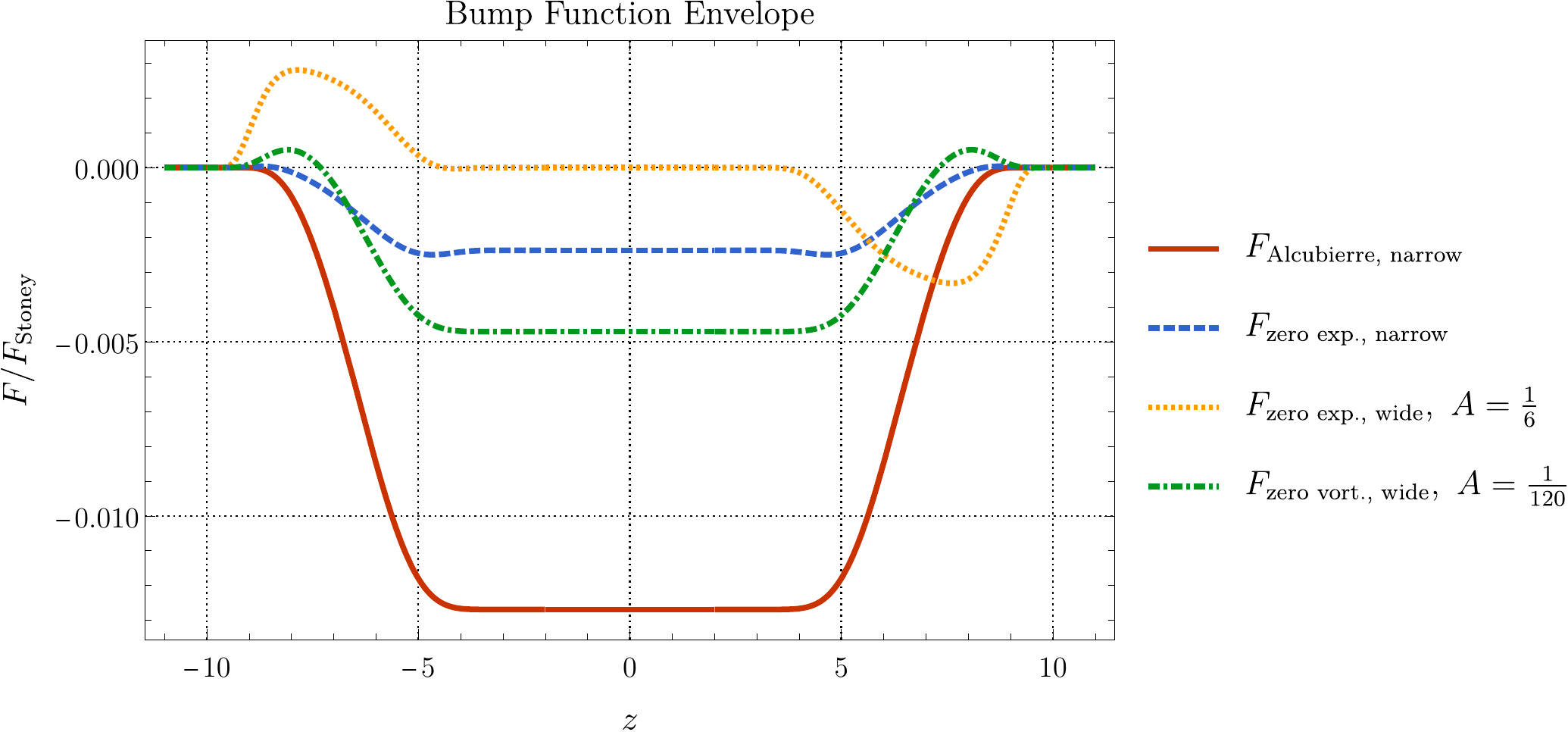}
 		\caption{}
 		\label{fig:BumpF1}    
 	\end{subfigure}\\
 	\begin{subfigure}[t]{\textwidth}
 		\includegraphics[width=.33\textwidth]{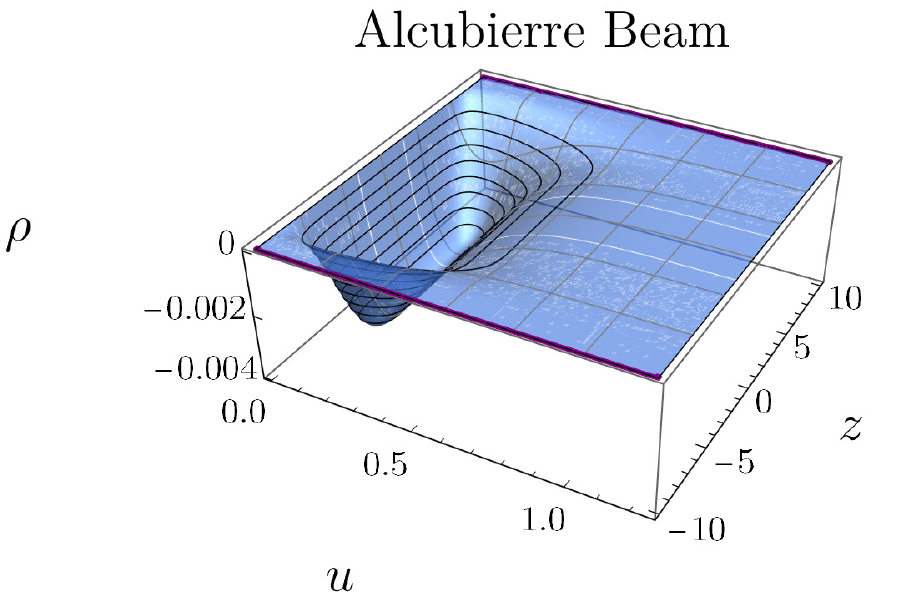}~
 		\includegraphics[width=.33\textwidth]{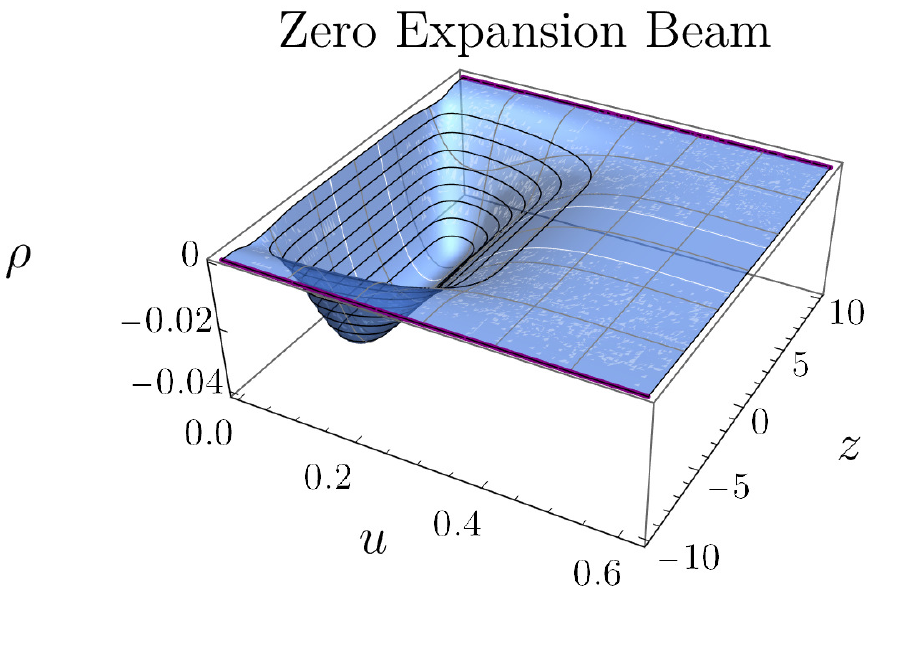}~
 		\includegraphics[width=.33\textwidth]{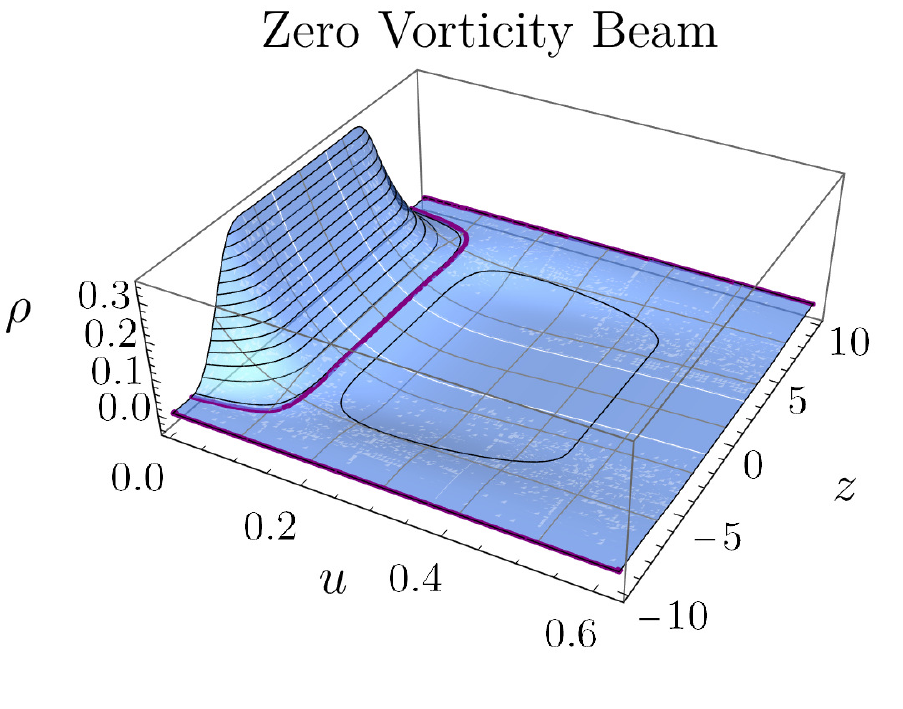}~
 		\caption{}
 		\label{fig:BumpRho1}    
 	\end{subfigure}
 	\caption{Forces (above) and energy densities (below) for beam profiles with smooth bump functions as described by equation~(\ref{E:bump}); the parameters in these pictures are $t=-1$, $a=2$, $b=10$, $A= {0.5}$, $B=C= {1.0}$ and $D= {1.0}$. The purple lines in the density plots indicate locations where the energy density is zero or indistinguishable from zero.}
 	\label{fig:Bump1}
 \end{figure}

\begin{figure}
 	\centering
 	\begin{subfigure}[t]{\textwidth}
 		\includegraphics[width=\textwidth]{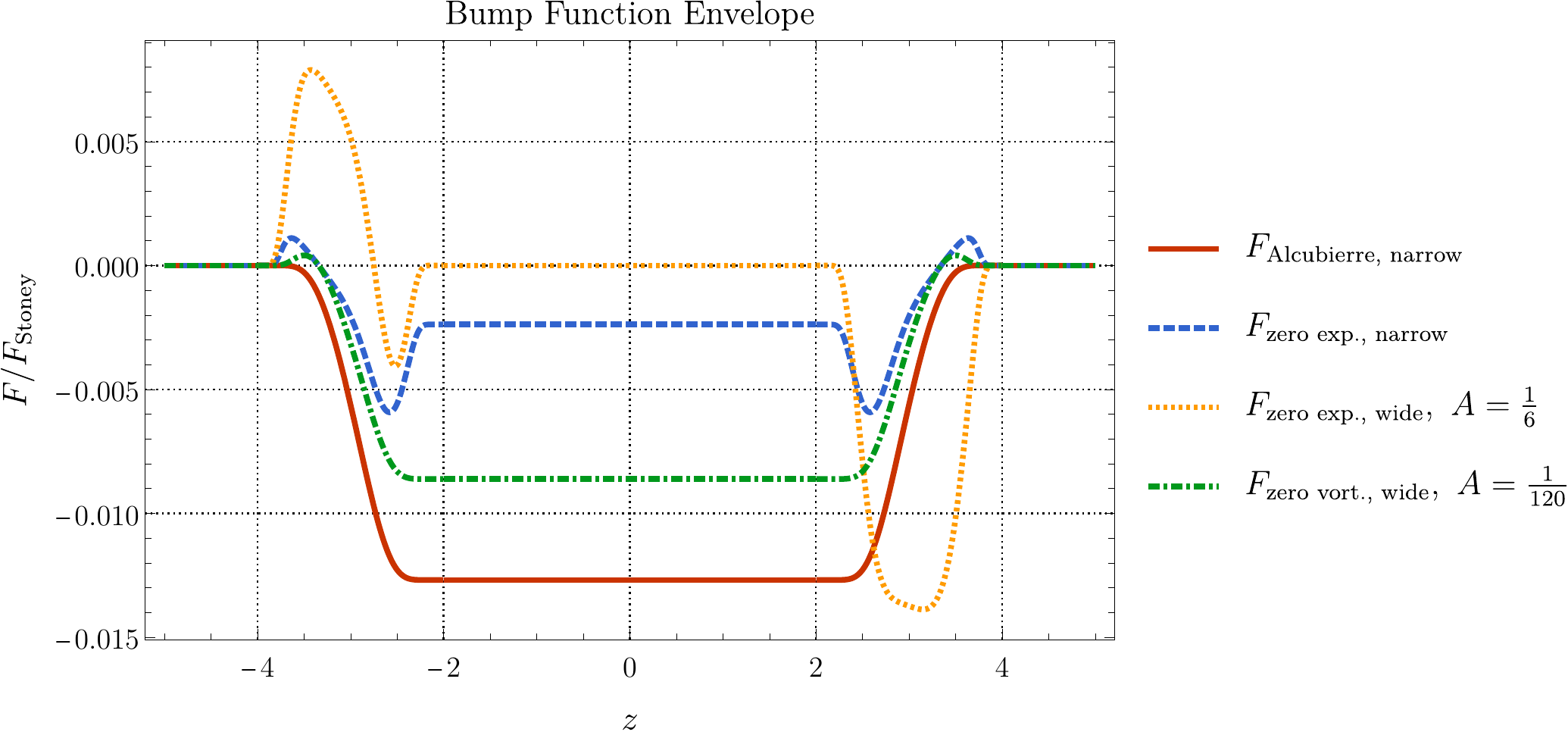}
 		\caption{}
 		\label{fig:BumpF2}    
 	\end{subfigure}\\
 	\begin{subfigure}[t]{\textwidth}
 		\includegraphics[width=.33\textwidth]{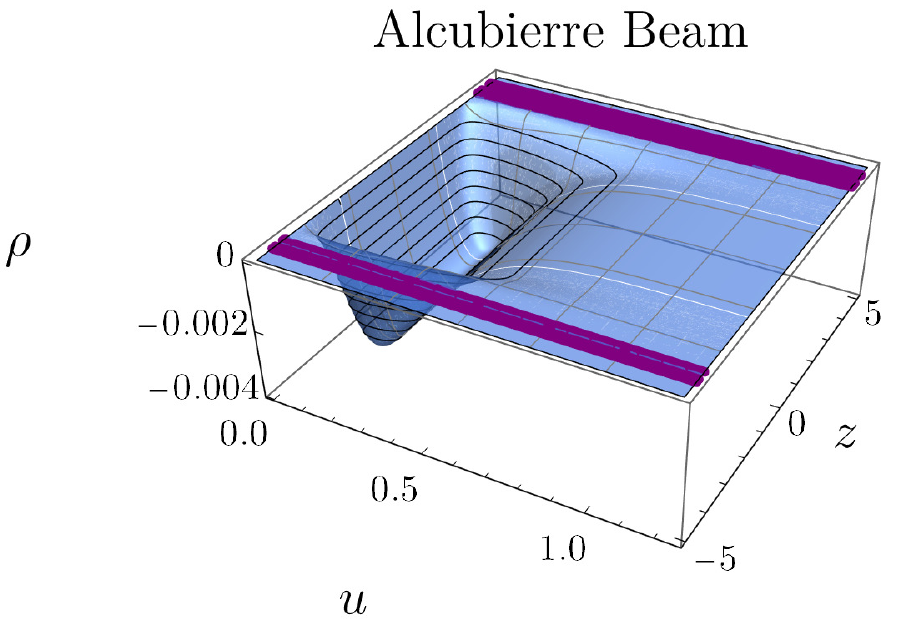}~
 		\includegraphics[width=.33\textwidth]{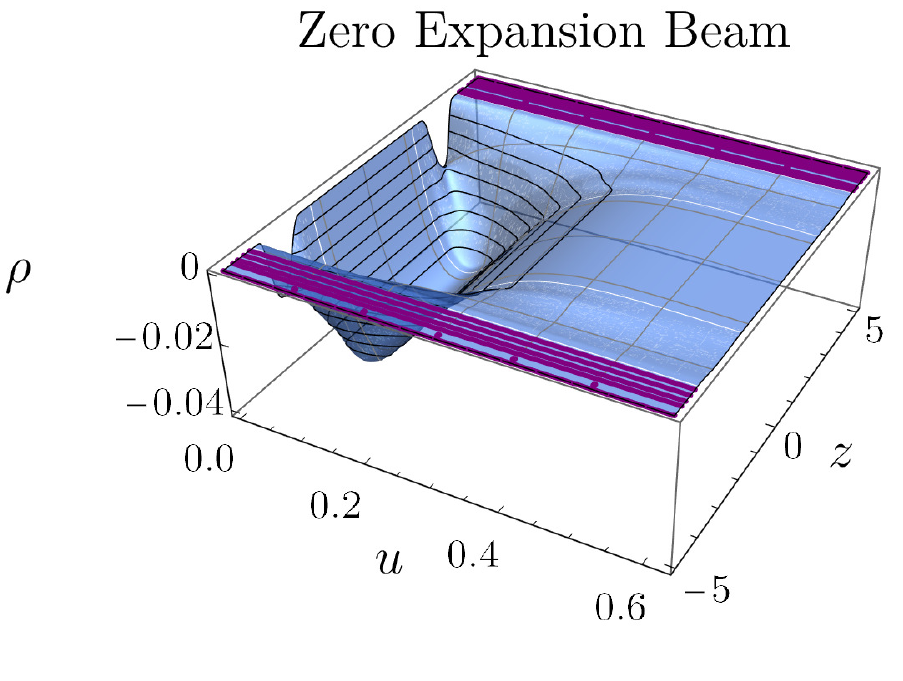}~
 		\includegraphics[width=.33\textwidth]{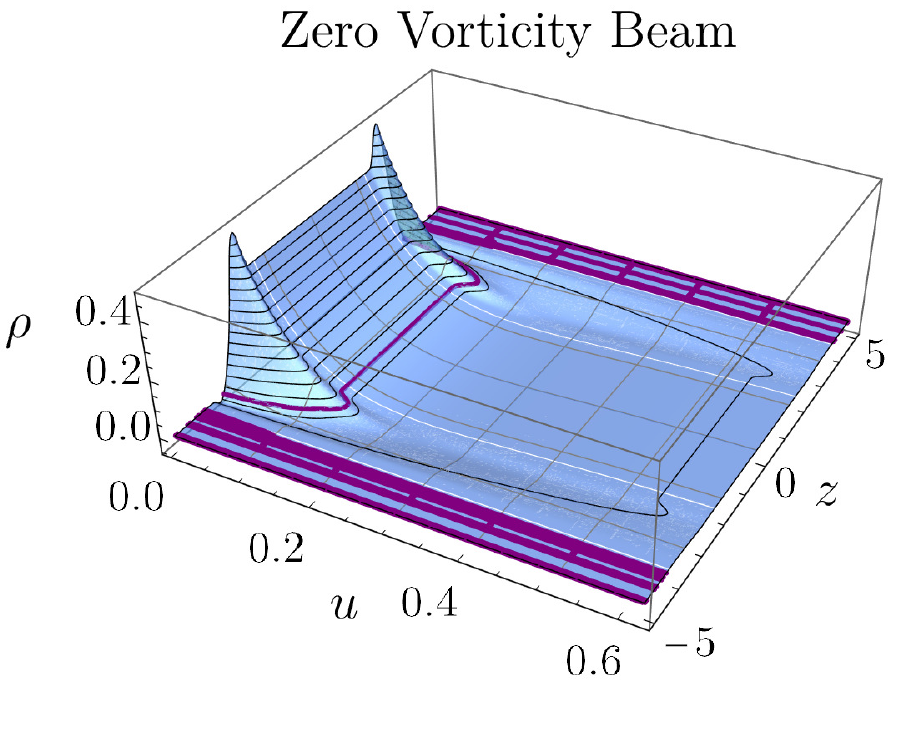}~
 		\caption{}
 		\label{fig:BumpRho2}    
 	\end{subfigure}
 	\caption{Forces (above) and energy densities (below) for beam profiles with smooth bump functions as described by equation~(\ref{E:bump}); the parameters in these pictures are $t=-1$, $a=2$, $b=4$, $A= {0.5}$, $B=C= {1.0}$ and $D= {1.0}$. The purple lines and bands in the density plots indicate locations where the energy density is zero or indistinguishable from zero.}
 	\label{fig:Bump2}
 \end{figure}

The algebra becomes arbitrarily involved in this case; for this reason we opt to only show our results and the functions we chose. The bump function used is
\begin{equation}\label{E:bump}
	f_{a,b}(z) \; \e^{-t^2/D^2},
\end{equation}
which, depending on the specific (special) case plotted, was used for $v$, $k$, or $\Phi$. The profile functions were again chosen to be the Gaussians, as described in section~\ref{sec:Gauss}, which also allows an easier comparison with the plots shown in figure~\ref{fig:Gauss}. In figure~\ref{fig:Bump1}, the parameters are $t=-1$, $a=2$, $b=10$, and $D=1$. In figure~\ref{fig:Bump2}, the parameters are $t=-1$, $a=2$, $b=4$, and $D=1$. Just this minor variation produced noticeable changes in the forces and energy density. The choice of $t$ can also produce significant differences, but this is not shown here, as it adds little to the discussion. 


 Again, note how non-trivial is the distribution of the energy density for the zero vorticity beams, which sums up to zero when integrated over any 3-spatial slice.	
It is also interesting to notice the different behaviour for distinct types of beam, varying from constant pull forces (\emph{e.g.} the Alcubierre case) up to elaborate push/pull behaviours (\emph{e.g.} wide zero-expansion). This reveals the great diversity of mechanisms one can create by varying the envelope functions only. Behaviour for different types of profile functions might possibly create yet other interesting scenarios, which we will leave for the enthusiastic reader.

 \clearpage
 Setting aside the issue of the magnitude of the Stoney force (which can be taken care of by an appropriately small pre-factor in our functions), we in particular like to draw attention to the force of the zero expansion beam in figure~\ref{fig:BumpF1}: A target positioned to the right at $z\approx 10$ would be accelerated to the left, then travel for a while at near constant velocity, before it is decelerated again. Sufficient fine-tuning thus allows for safe docking or boarding.

\FloatBarrier
\addtocontents{toc}{\protect\vspace{-5pt}}
\section{Discussion and conclusions}\label{S:conclusions}

In this article we have seen how to  analyze tractor/pressor/stressor beams within the framework of standard general relativity. The analysis was made based on modified warp drive spacetimes, by creating a ``beam like'' profile. A general case based on Natário's warp field was analyzed, followed by specific cases and examples.

As expected, we have seen that in this case, just like with warp drives and traversable wormholes, the violation of the NEC, and so of all the classical point-wise energy conditions, is unavoidable.
A closely related statement is still true even if one moves beyond Einstein gravity.
The key point is that it is ultimately the focussing properties of the tractor/pressor/stressor beams, warp fields and traversable wormholes that translate into convergence conditions~\cite{Raychaudhuri,Kar:2006,Dadhich:2005,Ehlers:2006,Borde:1987,vanVleck,Abreu:2010}, and thence into [effective] energy conditions.
Whenever you can rearrange the equations of motion in the form
\begin{equation}
G_{ab} = 8\pi [T_\mathrm{effective}]_{ab},
\end{equation}
then the effective energy-momentum tensor $ [T_\mathrm{effective}]_{ab}$ will consequently violate the NEC and so violate all the classical point-wise energy conditions. 

However, a significant question remains open: Are energy conditions truly fundamental physics? 
Probably not, (indeed, almost certainly not). But the energy conditions are certainly good diagnostics for unusual physics --- and, as we have seen, the physics of these tractor/pressor/stressor beams is  certainly extremely unusual --- comparable in weirdness to that of traversable wormholes and warp drives. This is not an absolute prohibition on tractor/pressor/stressor beams, but it is an invitation to think very carefully about the underlying physics.

\section*{Acknowledgements}
JS  acknowledges indirect financial support via the Marsden fund, 
administered by the Royal Society of New Zealand.

SS acknowledges financial support via OP RDE project No. CZ.02.2.69/0.0/0.0/18\_053/\-0016976 (International mobility of research), and the technical and administrative staff at the Charles University.

MV was directly supported by the Marsden Fund, via a grant administered by the Royal Society of New Zealand.

The authors also wish to thank Mariana Ruiz Villarreal for developing the two images used in figure~\ref{fig:GaussF}, and  releasing them into the public domain.\\
 (See \url{https://commons.wikimedia.org/wiki/File:Cow_cartoon_04.svg}\\ and \url{https://commons.wikimedia.org/wiki/File:Smiley_green_alien_nerdy.svg}.)

\bigskip
\bigskip
\hrule

\end{document}